\documentclass[twocolumn,showpacs,showkeys,preprintnumbers,amssymb,aps,superscriptaddress,prb]{revtex4-2}
\usepackage{graphicx}
\usepackage{dcolumn}
\usepackage{bm}
\usepackage{epsfig}
\usepackage{color}
\usepackage[breaklinks,colorlinks,bookmarks=false,citecolor=blue,linkcolor=red,urlcolor=blue]{hyperref}

\begin{document}

\title{Thermal first-order phase transitions, Ising critical points,  and reentrance in the
Ising-Heisenberg model on the diamond-decorated square lattice in a magnetic field}

\author{Jozef Stre\v{c}ka}
\affiliation{Department of Theoretical Physics and Astrophysics, Faculty of Science, P.~J. \v{S}af\'arik University, Park Angelinum 9, 04001 Ko\v{s}ice, Slovakia}
\author{Katar\'ina Karl'ov\'a}
\affiliation{Department of Theoretical Physics and Astrophysics, Faculty of Science, P.~J. \v{S}af\'arik University, Park Angelinum 9, 04001 Ko\v{s}ice, Slovakia}
\author{Taras Verkholyak}
\affiliation{Institute for Condensed Matter Physics, National Academy of Sciences of Ukraine, Svientsitskii Street 1, 790 11, L'viv, Ukraine}
\author{Nils Caci}
\affiliation{Institute for Theoretical Solid State Physics, JARA FIT, and JARA CSD, RWTH Aachen University, 52056 Aachen, Germany}
\author{Stefan Wessel}
\affiliation{Institute for Theoretical Solid State Physics, JARA FIT, and JARA CSD, RWTH Aachen University, 52056 Aachen, Germany}
\author{Andreas Honecker}
\affiliation{Laboratoire de Physique Th\'eorique et Mod\'elisation, CNRS UMR 8089, \\ CY Cergy Paris Universit\'e, 95000 Cergy-Pontoise, France}

\date{March 4, 2023}

\begin{abstract}
The thermal phase transitions of a spin-1/2 Ising-Heisenberg model on the diamond-decorated square lattice in a magnetic field are investigated using a decoration-iteration transformation and classical Monte Carlo simulations.
A generalized decoration-iteration transformation maps this model exactly onto an effective classical Ising model on the square lattice with temperature-dependent effective nearest-neighbor interactions and magnetic field strength. The effective field vanishes along a ground-state phase boundary of the original model, separating a  ferrimagnetic and a quantum monomer-dimer phase. At finite temperatures this phase boundary  gives rise to an exactly solvable surface of discontinuous (first-order) phase transitions, which terminates in a line of  Ising critical points. The existence of discontinuous reentrant phase transitions  within a narrow parameter regime is reported and explained in terms of the low-energy excitations from both phases. These exact results, obtained from the mapping  to the zero-field effective  Ising model  are 
corroborated by classical Monte Carlo simulations of the effective model. 
\end{abstract}


\maketitle

\section{Introduction}

Phase transitions and critical phenomena of spin systems in the presence of an external magnetic field remain a longstanding challenge for rigorous theoretical investigations~\cite{bax82}. For example, in spite of formidable efforts there does not exist a general exact solution to even the most basic  statistical lattice model, the two-dimensional spin-1/2 Ising model, in finite magnetic fields~\cite{bax82}. 
Exactly solvable  spin models displaying  thermal phase transitions and critical points in finite magnetic fields  therefore represent challenging research topics in their own right,  deserving  a great deal of attention.
An early example in this field is Fisher's `super-exchange antiferromagnet' \cite{fisher60a,fisher60b} where a magnetic field is applied to spins decorating the edges
of a square-lattice Ising model, but parameters nevertheless need to be fine-tuned to render this model exactly solvable.

From the experimental point of view, recent specific-heat measurements on the Shastry-Sutherland compound SrCu$_2$(BO$_3$)$_2$
\cite{sha81,miy03}
at zero magnetic field evidenced a remarkable line of discontinuous phase transitions in the pressure-temperature phase diagram~\cite{jim21}. The experimentally observed line of first-order transitions, terminating at a critical point that belongs to the universality class of the two-dimensional Ising model, was ascribed to a coexistence line of dimer-singlet and plaquette-singlet phases. Indeed, a theoretical modeling based on the notion of a pressure-tuned interaction ratio satisfactorily reproduced the measured specific-heat data~\cite{jim21}. Moreover, it turns out that the frustrated magnetic material SrCu$_2$(BO$_3$)$_2$ reveals in finite magnetic fields various quantum and thermal phase transitions in the pressure-temperature-field phase diagram, which currently represents a highly topical issue on account of their puzzling nature~\cite{jim21,tsu11,lee19,yan22,shi22,ima22}. It is worthwhile to remark that enigmatic finite-temperature phase transitions are not unique to SrCu$_2$(BO$_3$)$_2$, but may be encountered in various low-dimensional frustrated quantum materials (cf.\ the recent review \cite{kod22} and references cited therein).
Lines of first-order thermal phase transitions ending at  Ising critical points were recently reported also for the spin-1/2 Heisenberg square bilayer~\cite{sta18,web22b} and trilayer~\cite{web22}, as well as the spin-1/2 Heisenberg model on the diamond-decorated square lattice in finite magnetic fields~\cite{cac22}.

The spin-1/2 Heisenberg model on the diamond-decorated square lattice indeed provides a
prominent example of a dimer-based, geometrically frustrated quantum spin model that displays a great
diversity of  quantum ground states and phase transitions. Its zero-field ground-state
phase diagram  was comprehensively studied in Refs.~\onlinecite{mor16,hir16,hir17,hir18,hir20}, providing evidence for two unconventional ground-state phases, the monomer-dimer and
dimer-tetramer ground states, with  extensive residual entropies, in addition to the 
ferrimagnetic ground state expected according to the Lieb-Mattis theorem ~\cite{lie62}. Besides
the three aforementioned zero-field ground-state phases, in finite
magnetic fields the model exhibits a further spin-canted phase as well as a fully polarized paramagnetic phase~\cite{cac22}. The most interesting findings however concern the validation of peculiar
thermal phase transitions of the spin-1/2 Heisenberg model on the diamond-decorated square lattice by sign-problem-free quantum Monte Carlo simulations performed in a spin-dimer basis~\cite{cac22}. More specifically, a surface of first-order thermal transitions was identified, extending from the line of first-order quantum phase transitions that separates the monomer-dimer and the ferrimagnetic ground state phases. This wall of 
discontinuous thermal phase transitions eventually terminates in a line of Ising thermal critical points~\cite{cac22}, where the fluctuations in the dimer singlet (triplet) density from the monomer-dimer (ferrimagnetic) phase proliferate. 

The primary goal of the present article is to examine the aforementioned phenomena on more rigorous grounds. For that purpose we consider the analogous spin-1/2 Ising-Heisenberg model
on the diamond-decorated square lattice in a magnetic field, which provides  deeper insights into the aforementioned  thermal phase transitions and critical points via   exact analytical results. Here, the notion `Ising-Heisenberg model'  refers to an ensemble of interacting spins with Ising and Heisenberg interaction terms, which could be regarded more as an academic curiosity rather than a realistic model of some magnetic material. It nevertheless turns out that a few Ising-Heisenberg models adequately capture the magnetic behavior of specific magnetic materials, namely 
Cu(3-Clpy)$_2$(N$_3$)$_2$~\cite{str05}, 
[(CuL)$_2$Dy][Mo(CN)$_8$]~\cite{heu10,bel14}, 
[Fe(H$_2$O)(L)][Nb(CN)$_8$][Fe(L)]~\cite{sah12}, 
Dy(NO$_3$)(DMSO)$_2$Cu(opba)(DMSO)$_2$~\cite{str12,tor18}, 
\{Dy(hfac)$_2$(CH$_3$OH)\}$_2$\{Cu(dmg)(Hdmg)\}$_2$~\cite{str20,gal22} 
and [CuMn(L)][Fe(bpb)(CN)$_{2}$] ClO$_{4}$~\cite{sou20}. In the following, we will refer to a spin with only Ising interactions as an Ising spin, while all other spins of the Ising-Heisenberg model are referred to as Heisenberg spins. 

Thus far, the spin-1/2 Ising-Heisenberg model on the diamond-decorated square lattice was only exactly solved for zero magnetic field, which provided proof for a line of continuous (second-order) quantum phase transitions separating a spontaneously long-range ordered phase from a disordered one~\cite{str06}. The magnetic behavior of this mixed classical-quantum spin model in nonzero magnetic fields has not been dealt with previously and will be the subject  of the present article, which is organized as follows. In Sec.~\ref{model} we  introduce the spin-1/2 Ising-Heisenberg model on the diamond-decorated square lattice and explain the  basic steps of the  methods applied for its solution. The most interesting results regarding the phase diagram, magnetization curves and the order parameter are then discussed in Sec.~\ref{results}. Finally, the paper ends with a summary and perspectives in Sec.~\ref{conclusion}. 

\section{Model and method}
\label{model}

\begin{figure}[t!]
\begin{center}
\includegraphics[width=0.6\columnwidth]{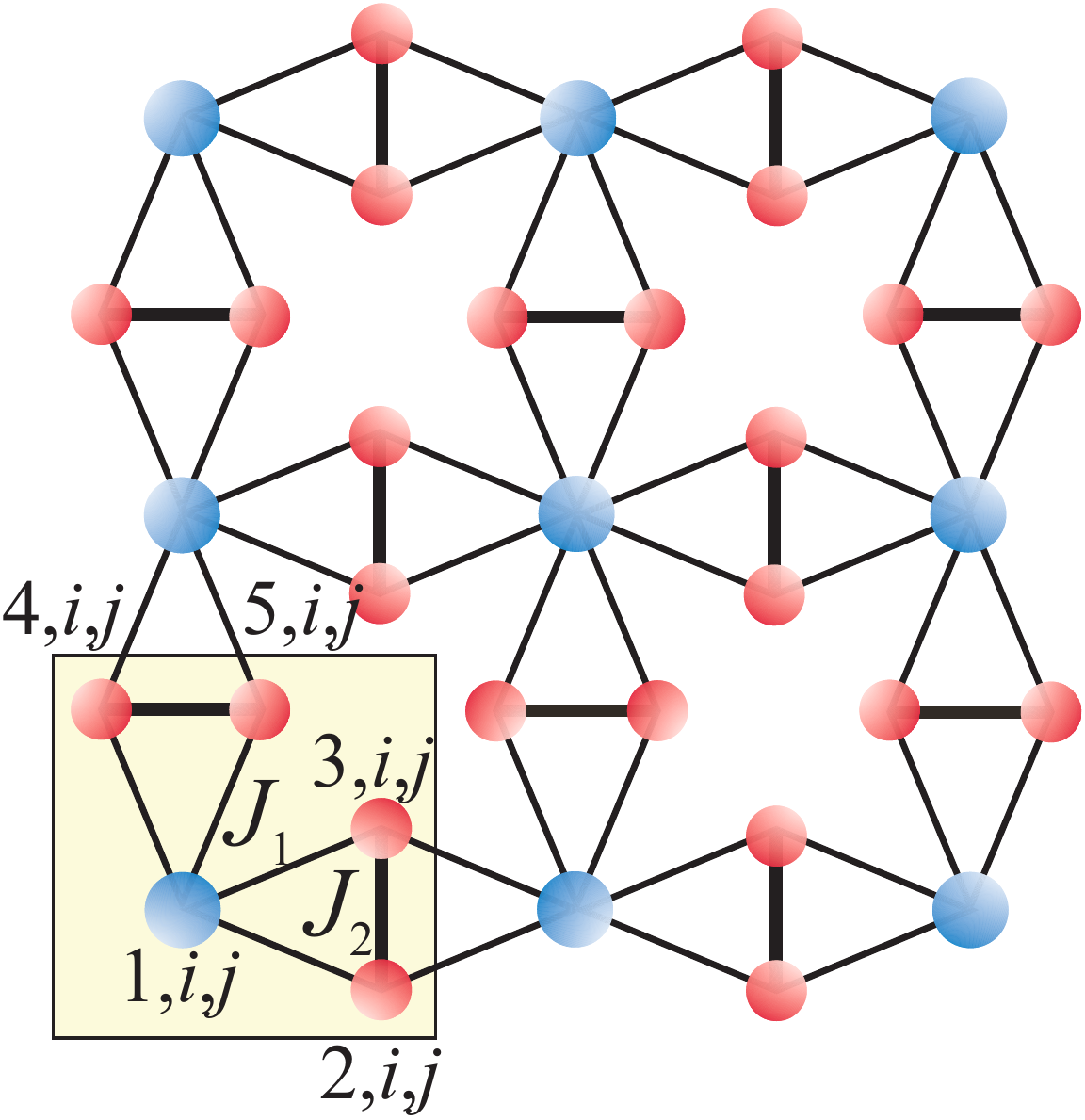}
\end{center}
\vspace{-0.5cm}
\caption{Illustration of  the diamond-decorated square lattice. Large blue (small red) circles  denote  Ising (Heisenberg)  spins. The unit cell, containing five spins, is also indicated.}
\label{lattice}       
\end{figure}

The spin-1/2 Ising-Heisenberg model on the diamond-decorated square lattice, schematically illustrated in Fig.~\ref{lattice}, is  given by the Hamiltonian
\begin{eqnarray}
\hat{H} &=& J_1 \sum_{i=1}^{L} \sum_{j=1}^{L} 
\Bigl[ \left(\hat{{S}}_{1,i,j}^z + \hat{{S}}_{1,i+1,j}^z \right) 
\left(\hat{{S}}^z_{2,i,j} + \hat{{S}}^z_{3,i,j} \right) \nonumber \\ 
&& \qquad \qquad + \left(\hat{{S}}_{1,i,j}^z + \hat{{S}}_{1,i,j+1}^z \right) 
\left(\hat{{S}}^z_{4,i,j}+\hat{{S}}^z_{5,i,j}\right) \Bigr] \nonumber \\
&&+ J_2 \sum_{i=1}^{L} \sum_{j=1}^{L} 
\left(\hat{\bf {S}}_{2,i,j}\cdot\hat{\bf {S}}_{3,i,j} 
+ \hat{\bf {S}}_{4,i,j}\cdot\hat{\bf {S}}_{5,i,j}\right) \nonumber \\  
&&- h \sum_{k=1}^{5} \sum_{i=1}^{L} \sum_{j=1}^{L} \hat{S}_{k,i,j}^z. 
\label{eq:ham}
\end{eqnarray}
Here, a spin-1/2 operator $\hat{\bf S}_{k,i,j} \equiv (\hat{S}_{k,i,j}^{x}, \hat{S}_{k,i,j}^{y}, \hat{S}_{k,i,j}^{z})$ is assigned to each site of the diamond-decorated square lattice, where the former subscript $k$ determines the site position within the unit cell and the latter two subscripts $i$ and $j$ specify the position (in terms of row and column) of the unit cell itself.
According to the Hamiltonian (\ref{eq:ham}), the nodal sites ($k=1$) of the diamond-decorated square lattice shown in Fig.~\ref{lattice} as large blue circles are occupied by Ising spins, while decorating sites ($k=2,3,4,5$) schematically visualized as small red circles are occupied by Heisenberg spins. The coupling constant $J_1$  takes into account the Ising-type exchange interaction between  nearest-neighbor Ising and Heisenberg spins, while the coupling constant $J_2$ quantifies  the isotropic exchange interaction between the nearest-neighbor Heisenberg spins. The last term $h$ accounts for the Zeeman energy of both Heisenberg and Ising spins in the external magnetic field.

For further convenience, the total Hamiltonian~(\ref{eq:ham}) of the spin-1/2 Ising-Heisenberg model on the diamond-decorated square lattice can be rewritten as a sum of cluster Hamiltonians:
\begin{eqnarray}
\hat{H} = \displaystyle \sum_{i=1}^L \sum_{j=1}^L (\hat{H}_{i,j}^h + \hat{H}_{i,j}^v), 
\end{eqnarray}
where each cluster Hamiltonian $\hat{H}_{i,j}^h$ ($\hat{H}_{i,j}^v$) includes all interaction terms of a diamond spin cluster, containing one horizontal (vertical) Heisenberg dimer, as schematically shown in the left part of Fig.~\ref{local}. These read explicitly:
\begin{eqnarray}
\hat{H}_{i,j}^h &=& J_1 (\hat{{S}}_{1,i,j}^z \!+\! \hat{{S}}_{1,i+1,j}^z)(\hat{{S}}^z_{2,i,j}\!+\!\hat{{S}}^z_{3,i,j}) 
+ J_2 \hat{\bf {S}}_{2,i,j} \!\cdot\! \hat{\bf {S}}_{3,i,j}\nonumber \\
 &&- h (\hat{S}_{2,i,j}^z \!+\! \hat{S}_{3,i,j}^z) - \frac{h}{4} (\hat{{S}}_{1,i,j}^z \!+\! \hat{{S}}_{1,i+1,j}^z), \nonumber \\
\hat{H}_{i,j}^v &=& J_1 (\hat{{S}}_{1,i,j}^z \!+\! \hat{{S}}_{1,i,j+1}^z)(\hat{{S}}^z_{4,i,j}\!+\!\hat{{S}}^z_{5,i,j}) 
+ J_2 \hat{\bf {S}}_{4,i,j} \!\cdot\! \hat{\bf {S}}_{5,i,j}\nonumber \\
&&- h (\hat{S}_{4,i,j}^z \!+\! \hat{S}_{5,i,j}^z) - \frac{h}{4} (\hat{{S}}_{1,i,j}^z \!+\! \hat{{S}}_{1,i,j+1}^z). 
\label{eq:hamdsc}
\end{eqnarray}
Note that the factor $1/4$ in the Zeeman term of the Ising spins ensures the proper counting of this interaction term, which is symmetrically split among four cluster Hamiltonians, involving one and the same Ising spin. The decomposition of the total Hamiltonian (\ref{eq:ham}) into a set of commuting cluster Hamiltonians (\ref{eq:hamdsc}) allows one to partially factorize the total partition function into a product of the cluster partition functions:
\begin{figure}[t!]
\begin{center}
\includegraphics[width=0.8\columnwidth]{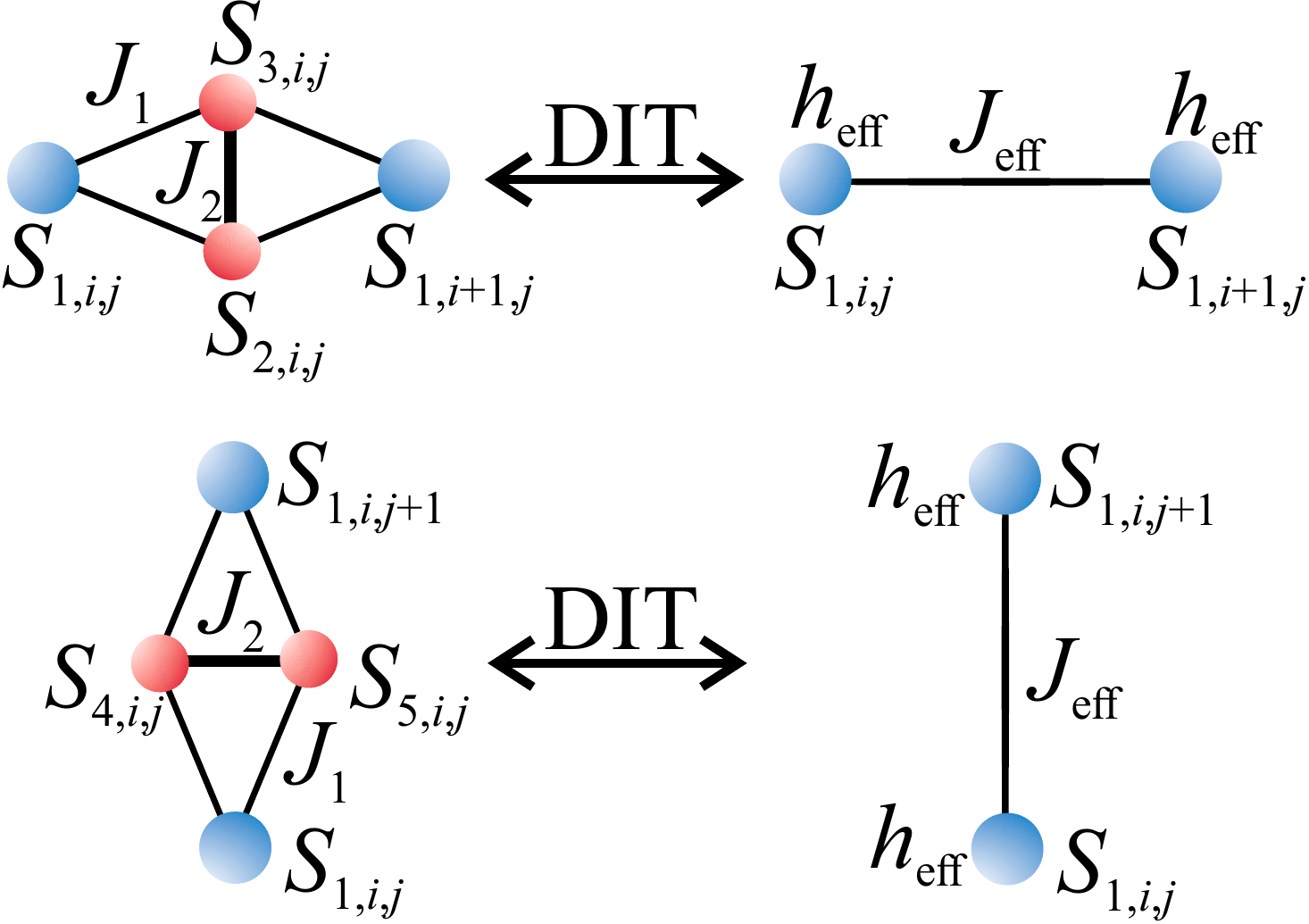}
\end{center}
\vspace{-0.5cm}
\caption{Illustration of the decoration-iteration transformations (DIT)  (\ref{dith}) and (\ref{ditv}) applied locally to a horizontal and vertical diamond spin cluster, respectively.}
\label{local}       
\end{figure}
\begin{eqnarray}
{Z} = \sum_{\{S_{1,i,j}^z\}} \prod_{i=1}^L \prod_{j=1}^L 
&&[\mbox{Tr}_{2,i,j} \mbox{Tr}_{3,i,j} \exp(-\beta \hat{H}_{i,j}^h)]\nonumber\\
&&\times [\mbox{Tr}_{4,i,j} \mbox{Tr}_{5,i,j} \exp(-\beta \hat{H}_{i,j}^v)].
\quad
\label{pff}
\end{eqnarray}
Formula (\ref{pff}) is consistent with the fact that the spin degrees of freedom of the Heisenberg dimers can be traced out independently of each other. A straightforward diagonalization of the cluster Hamiltonians (\ref{eq:hamdsc}) within the spin-dimer basis provides the effective Boltzmann weights, which can be substituted through the generalized decoration-iteration transformation \cite{fis59,roj09,str10}. An explicit form of the decoration-iteration transformation for the horizontal and vertical diamond spin clusters is given by the following formulas:
\onecolumngrid
\begin{eqnarray}
\mbox{Tr}_{2,i,j} \mbox{Tr}_{3,i,j} \exp(-\beta \hat{H}_{i,j}^h) &=& 
\exp\!\left[{ \frac{3 \beta J_2}{4}} + \frac{\beta h}{4} \left(S_{1,i,j}^z \!+\! S_{1,i+1,j}^z\right)\right]
\left\{1 + {\rm e}^{-\beta J_2} \! \left[1 + 2 \cosh \left( \beta J_1 (S_{1,i,j}^z \!+\! S_{1,i+1,j}^z) - \beta h \right) \right] \right\}
\nonumber \\
&=& A \exp\left[\beta J_{\rm eff} S_{1,i,j}^z S_{1,i+1,j}^z 
+ \frac{\beta h_{\rm eff}}{4} \left(S_{1,i,j}^z \!+\! S_{1,i+1,j}^z\right)\right]\!\!,
\label{dith}
\end{eqnarray}
and
\begin{eqnarray}
\mbox{Tr}_{4,i,j} \mbox{Tr}_{5,i,j} \exp(-\beta \hat{H}_{i,j}^v) &=& 
\exp\!\left[{ \frac{3 \beta J_2}{4}} + \frac{\beta h}{4} \left(S_{1,i,j}^z \!+\! S_{1,i,j+1}^z\right)\right]
\left\{1 + {\rm e}^{-\beta J_2} \! \left[1 + 2 \cosh \left( \beta J_1 (S_{1,i,j}^z \!+\! S_{1,i,j+1}^z) - \beta h \right) \right] \right\}
\nonumber \\
&=& A \exp\left[\beta J_{\rm eff} S_{1,i,j}^z S_{1,i,j+1}^z 
+ \frac{\beta h_{\rm eff}}{4} \left(S_{1,i,j}^z \!+\! S_{1,i,j+1}^z\right)\right]\!\!.
\label{ditv}
\end{eqnarray} 
\twocolumngrid
\noindent
The physical meaning of the decoration-iteration transformations (\ref{dith}) and (\ref{ditv}) lies in replacing the Boltzmann weights related to the diamond spin cluster by an equivalent expression, which exclusively depends only on two Ising spins.
It directly follows from Eqs.\ (\ref{dith}) and (\ref{ditv}) that the Heisenberg spin dimer and its associated interaction terms can be replaced by the effective interaction $J_{\rm eff}$ and effective field $h_{\rm eff}$ ascribed to two enclosing Ising spins (see Fig.\ \ref{local} for a schematic representation of this mapping). 
This is actually reminiscent of the procedure used in Refs.~\onlinecite{fisher60a,fisher60b}, except that here we trace out a Heisenberg dimer instead of a single Ising spin per edge of the effective square lattice.

Note that one gets only three independent equations from the decoration-iteration transformations (\ref{dith}) and (\ref{ditv}) by considering all four possible combinations of two Ising spins, which unambiguously determine the  transformation parameters as
$A$, $J_{\rm eff}$ and $h_{\rm eff}$:
\begin{eqnarray}
A &=& { {\rm e}^{\frac{3}{4}\beta J_2}} \left(V_1 V_2 V_3^2 \right)^{\frac{1}{4}}, \label{mpa} \\
\beta J_{\rm eff} &=& \ln \left(\frac{V_1 V_2}{V_3^2} \right), \label{mpj} \\
\beta h_{\rm eff} &=& \beta h + 2 \ln \left(\frac{V_1}{V_2} \right), \label{mph} 
\end{eqnarray}
where
\begin{eqnarray}
V_1 &=& 1 + {\rm e}^{-\beta J_2} \left[1 + 2 \cosh \left(\beta J_1 - \beta h \right) \right], \nonumber \\
V_2 &=& 1 + {\rm e}^{-\beta J_2} \left[1 + 2 \cosh \left(\beta J_1 + \beta h \right) \right], \nonumber \\
V_3 &=& 1 + {\rm e}^{-\beta J_2} \left[1 + 2 \cosh \left(\beta h \right) \right]. \label{v1v3}
\end{eqnarray}
Using the formulas (\ref{dith}) and (\ref{ditv}) in the factorized form of the partition function (\ref{pff}) establishes an exact mapping  between the partition functions of the spin-1/2 Ising-Heisenberg model on the diamond-decorated square lattice and the effective spin-1/2 Ising model on a square lattice:
\begin{figure}[t!]
\begin{center}
\includegraphics[width=\columnwidth]{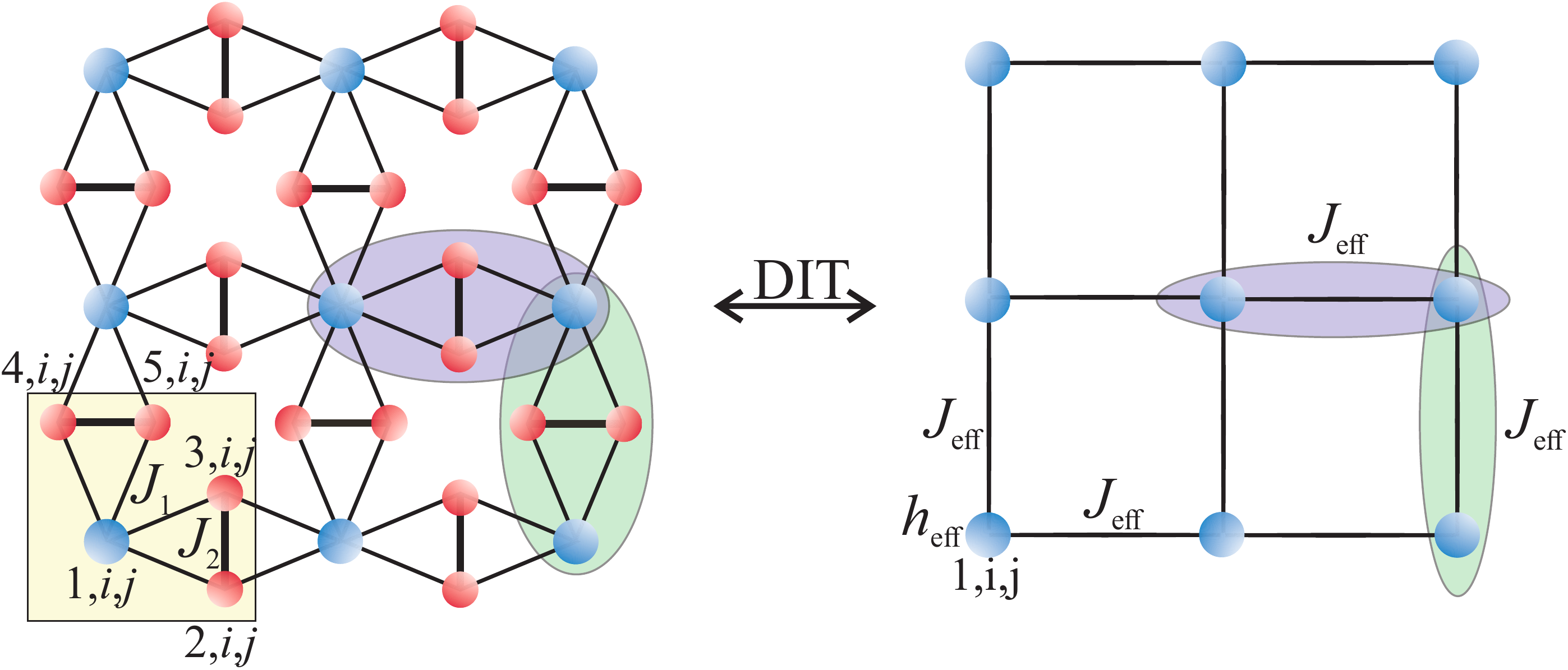}
\end{center}
\vspace{-0.5cm}
\caption{Mapping between the spin-1/2 Ising-Heisenberg diamond-decorated square lattice and the effective spin-1/2  square-lattice Ising model obtained by applying the decoration-iteration transformations  (DIT) (\ref{dith}) and (\ref{ditv}) to all horizontal and vertical diamond spin clusters.}
\label{global}       
\end{figure}
\begin{eqnarray}
{Z} (\beta, J_1, J_2, h) =  A^{2N} {Z}_{\rm eff} (\beta, J_{\rm eff}, h_{\rm eff}), 
\label{zm}
\end{eqnarray}  
which is defined through an effective Hamiltonian involving  temperature-dependent nearest-neighbor interactions $J_{\rm eff}$ and magnetic field $h_{\rm eff}$ (see Fig.\ \ref{global} for an illustration of the established mapping):
\begin{eqnarray}
{H}_{\rm eff} = &-& J_{\rm eff} \sum_{i=1}^L \sum_{j=1}^L (S_{1,i,j}^z S_{1,i+1,j}^z + S_{1,i,j}^z S_{1,i,j+1}^z) \nonumber \\
                     &-& h_{\rm eff} \sum_{i=1}^L \sum_{j=1}^L S_{1,i,j}^z. 
\label{hef}										
\end{eqnarray}
Since the coefficient $A$ in Eq.~(\ref{zm}) is a regular function  (see Eq.~(\ref{mpa})), $Z$ inherits the singular structure from ${Z}_{\rm eff}$ such that any emerging continuous thermal phase transition will be in the Ising universality class.

{Local observables can be calculated from the  correspondence between the spin-1/2 Ising-Heisenberg diamond-decorated square lattice model and the effective Ising model as follows: The local magnetization of the original Ising spins and the  correlation function among nearest-neighbor Ising spins follow from the  theorems by Barry \textit{et al}., \cite{bar88,kha90,bar91,bar95}
\begin{eqnarray}
m_{\rm I} &\equiv& \langle \hat{S}_{1,i,j}^z \rangle = \langle S_{1,i,j}^z \rangle_{\rm eff} \equiv m_{\rm eff} \nonumber \\
\varepsilon_{\rm I} &\equiv& \langle \hat{S}_{1,i,j}^z \hat{S}_{1,i+1,j}^z \rangle = \langle S_{1,i,j}^z S_{1,i+1,j}^z \rangle_{\rm eff} \equiv \varepsilon_{\rm eff}, 
\label{miei}										
\end{eqnarray}
where  $\langle \cdots \rangle$ and $\langle \cdots \rangle_{\rm eff}$ denote the standard canonical ensemble average within the spin-1/2 Ising-Heisenberg diamond-decorated square lattice Hamiltonian (\ref{eq:ham}) and the effective Ising model Hamiltonian (\ref{hef}), respectively. Hence, the local magnetization $m_{\rm I}$ and the nearest-neighbor correlation function $\varepsilon_{\rm I}$ are directly equal to the quantities $m_{\rm eff}$ and $\varepsilon_{\rm eff}$ of the  effective Ising model with effective interaction $J_{\rm eff}$ and  effective field $h_{\rm eff}$.
In addition, the mapping (\ref{zm}) between the partition functions in turn provides an explicit formula for the free energy of the spin-1/2 Ising-Heisenberg diamond-decorated square lattice model $F = - k_{\rm B} T \ln Z$, from which one may calculate the total magnetization accordingly, 
\begin{eqnarray}
M_{\rm T} = - \frac{\partial F}{\partial h} = 2N \frac{\partial \ln A}{\partial (\beta h)} + \frac{\partial \ln Z_{\rm eff}}{\partial (\beta h)}. 
\label{MT}										
\end{eqnarray}
The total magnetization $M_{\rm T} = N m_{\rm I} + 4 N m_{\rm H}$ can alternatively be  expressed in terms of the single-site magnetization of the Ising spins $m_{\rm I} \equiv \langle \hat{S}_{1,i,j}^z \rangle$ and the single-site magnetization of the Heisenberg spins $m_{\rm H} \equiv \langle \hat{S}_{k,i,j}^z \rangle$ ($k=2, \ldots, 5$), so that the final formula (\ref{MT}) for the total magnetization allows for a straightforward derivation of the single-site magnetization of the Heisenberg spins upon subtracting the relevant contribution from the Ising spins:
\begin{eqnarray}
m_{\rm H} &\equiv& \langle \hat{S}_{k,i,j}^z \rangle = \frac{1}{8} \! \left(\frac{W_1}{V_1} \!+\! \frac{W_2}{V_2} \!+\! 2 \frac{W_3}{V_3} \right) 
+ \frac{m_{\rm I}}{2} 
\left(\frac{W_1}{V_1} \!-\! \frac{W_2}{V_2} \right) 
\nonumber \\
&&+ \frac{\varepsilon_{\rm I}}{2} 
\left(\frac{W_1}{V_1} \!+\! \frac{W_2}{V_2} \!-\! 2 \frac{W_3}{V_3} \right), \,\, (k = 2, \ldots, 5),
\label{mh}										
\end{eqnarray}
where we have introduced the following coefficients $W_1 = - 2 {\rm e}^{-\beta J_2} \sinh(\beta J_1 - \beta h)$, $W_2 = 2 {\rm e}^{-\beta J_2} \sinh(\beta J_1 + \beta h)$, and 
$W_3 = 2 {\rm e}^{-\beta J_2} \sinh(\beta h)$. 

Another local quantity that is of  importance is the density of the singlet-dimer states of the Heisenberg spin pairs. The density of these singlets can be evaluated from the nearest-neighbor correlation function of the Heisenberg spins according to 
\begin{eqnarray}
n = \frac{1}{4} - \langle \hat{\bf {S}}_{2k,i,j}\cdot\hat{\bf {S}}_{2k+1,i,j} \rangle  = \frac{1}{4} + \frac{1}{2N} \frac{\partial \ln Z}{\partial (\beta J_2)}.  
\label{dens}										
\end{eqnarray}
After performing the respective differentiation in Eq.~(\ref{dens}), the density of singlets can be expressed in terms of the single-site magnetization of the Ising spins $m_{\rm I}$ and the nearest-neighbor correlation function 
$\varepsilon_{\rm I}$ of the Ising spins:
\begin{eqnarray}
n &=& \frac{1}{4} \! \left(\frac{1}{V_1} \!+\! \frac{1}{V_2} \!+\! \frac{2}{V_3} \right) 
+ m_{\rm I} \! \left(\frac{1}{V_1} \!-\! \frac{1}{V_2} \right) 
\nonumber \\
&&+ \varepsilon_{\rm I} \! \left(\frac{1}{V_1} \!+\! \frac{1}{V_2} \!-\! \frac{2}{V_3} \right). 
\label{denss}										
\end{eqnarray} 
}

Recall that the zero-field spin-1/2 Ising model on a square lattice is exactly solvable from Onsager's ingenious exact solution \cite{ons44}. Thanks to the exact mapping relation (\ref{zm}) between both partition functions, the spin-1/2 Ising-Heisenberg model on the diamond-decorated square lattice in a magnetic field thus also becomes exactly soluble in the particular parameter subspace where the effective field vanishes, $h_{\rm eff} = 0$. From Eq.\ (\ref{mph}), this condition yields: 
\begin{eqnarray}
\exp\left(\frac{\beta h}{2}\right) = \frac{1 + {\rm e}^{-\beta J_2} \left[1 + 2 \cosh \left(\beta J_1 \!+\! \beta h \right) \right]}
{1 + {\rm e}^{-\beta J_2} \left[1 + 2 \cosh \left(\beta J_1 \!-\! \beta h \right) \right]}.
\label{zef}
\end{eqnarray}
It  turns out that the transcendent equation (\ref{zef}), ensuring a vanishing  effective field $h_{\rm eff} = 0$, also has a nontrivial solution for finite magnetic field(s) $h \neq 0$ and a given temperature apart from the trivial solution $h=0$ that is valid for any temperature. One of the intriguing consequences of Onsager's exact solution~\cite{ons44} is that the spin-1/2 Ising  model on the square lattice in zero (effective) field shows a continuous phase transition at the critical temperature $\beta_c J_{\rm eff} = J_{\rm eff}/(k_{\rm B} T_{\rm c}) = 2 \ln (1 + \sqrt{2})$, which in turn implies the existence of analogous continuous phase transitions within the universality class of the two-dimensional Ising model for the original  Ising-Heisenberg model whenever the following critical condition is met:
\begin{eqnarray}
\exp\left(\!\frac{\beta_c h}{4}\!\right) \! (1\!+\!\sqrt{2}) = 
\frac{1 + {\rm e}^{-\beta_c J_2} \left[1 + 2 \cosh \left(\beta_c (J_1 \!+\! h) \right) \right]}
{1 + {\rm e}^{-\beta_c J_2} \left[1 + 2 \cosh \left(\beta_c h \right) \right]}.  \nonumber \\
\label{critical}
\end{eqnarray}
The spin-1/2 Ising model on a square lattice in nonzero (effective) magnetic field is not exactly solvable and one must therefore resort to  numerical methods if one intends to obtain further quantitative results for the spin-1/2 Ising-Heisenberg model on the diamond-decorated square lattice from the mapping relation (\ref{zm}) based on  the effective spin-1/2 Ising model on the square lattice. For this purpose, we performed classical Monte Carlo simulations of the effective spin-1/2 Ising model on the square lattice using the standard Metropolis algorithm of an open source software from the Algorithms and Libraries for Physics Simulations (ALPS) project \cite{alps} for finite-size lattices with linear lattice sizes up to $L=120$ and a total number of  up to 
$8 \times 10^{5}$ Monte Carlo steps. 

\section{Results and discussion}
\label{results}

In the following,  we will proceed to a discussion of the most interesting results for the ground-state and finite-temperature phase diagrams, magnetization curves, and singlet density of the spin-1/2 Ising-Heisenberg model on the diamond-decorated square lattice in the presence of an external magnetic field.

\subsection{Ground-state phase diagram}

\begin{figure}[t!]
\begin{center}
\includegraphics[width=\columnwidth]{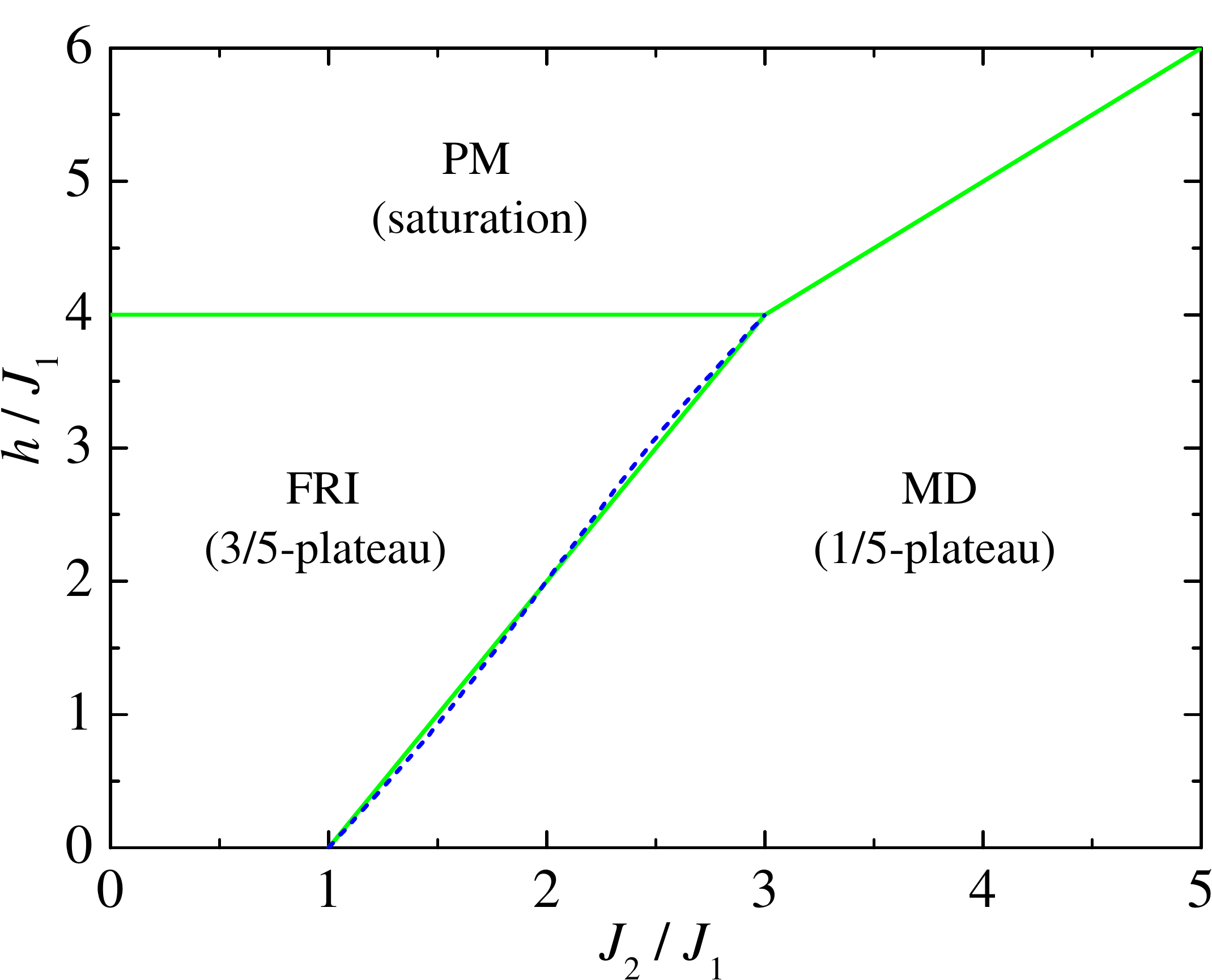}
\end{center}
\vspace{-0.5cm}
\caption{Ground-state phase diagram of the spin-1/2 Ising-Heisenberg model on the diamond-decorated square lattice in the $J_2/J_1-h/J_1$ plane. The notation for the various ground states is as follows: FRI -- classical ferrimagnetic phase, MD -- quantum monomer-dimer phase and PM -- saturated paramagnetic phase. The dashed blue line shows the projection of the critical condition (\ref{critical}) onto the $J_2/J_1-h/J_1$ plane.}
\label{gspd}       
\end{figure}

We begin our discussion by presenting the ground-state phase diagram, shown in Fig.~\ref{gspd} in the $J_2/J_1-h/J_1$ plane.
All ground states can be derived from the lowest-energy eigenstates of the diamond spin clusters given by the commuting local Hamiltonians (\ref{eq:hamdsc}). Using this procedure, we obtain three different ground state phases: the saturated paramagnetic (PM) phase, the classical ferrimagnetic (FRI) phase and the quantum monomer-dimer (MD) phase (cf.\ the detailed definitions below). By comparing the respective ground-state energies one obtains exact formulas for the first-order phase boundaries:
\begin{eqnarray}
\label{h_boundary}
&& h_\mathrm{MD-FRI} = 2(J_2 - J_1), \label{Eq:hMDFRI}\\
&& h_\mathrm{MD-PM} = J_2 + J_1, \\
&& h_\mathrm{FRI-PM} = 4J_1.
\end{eqnarray}
All three discontinuous phase-transition lines meet at a triple point 
 $J_2/J_1 = 3$, $h/J_1 = 4$.
\begin{figure}[t!]
\begin{center}
\includegraphics[width=\columnwidth]{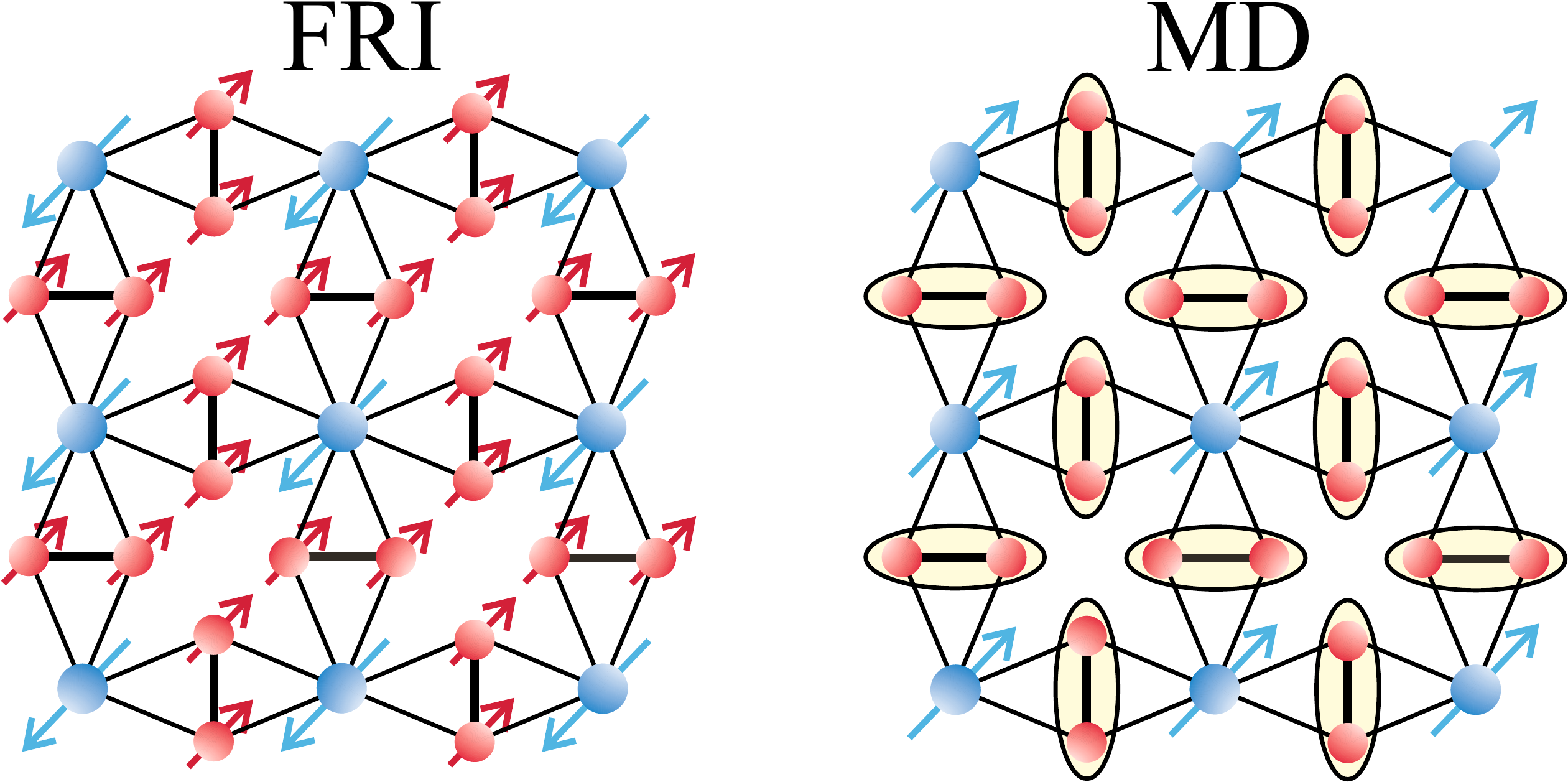}
\end{center}
\vspace{-0.5cm}
\caption{Schematic representation of the classical ferrimagnetic (FRI) phase and the quantum monomer-dimer (MD) phase. An oval represents a singlet-dimer state.}
\label{phases}       
\end{figure}
At sufficiently large magnetic fields $h>h_\mathrm{FRI-PM}$ for $J_2<3J_1$ and $h>h_\mathrm{MD-PM}$ for $J_2>3J_1$, the ground state is in the PM phase with all spins polarized along the magnetic-field direction: 
\begin{eqnarray}
|{\rm PM} \rangle = \prod_{i,j=1}^L \! |\!\!\uparrow_{1,i,j}\rangle \!\otimes\! |\!\!\uparrow_{2,i,j}\uparrow_{3,i,j}\rangle \!\otimes\! |\!\!\uparrow_{4,i,j}\uparrow_{5,i,j}\rangle.
\end{eqnarray}
Apart from this rather trivial phase two additional ground states schematically illustrated in Fig.\ \ref{phases} emerge at low enough magnetic fields. The first ground state can be identified as the FRI phase with all Heisenberg spins fully polarized  and all Ising spins aligned opposite to the magnetic field: 
\begin{eqnarray}
|{\rm FRI} \rangle = \prod_{i,j=1}^L \! |\!\!\downarrow_{1,i,j}\rangle \!\otimes\! |\!\!\uparrow_{2,i,j}\uparrow_{3,i,j}\rangle \!\otimes\! |\!\!\uparrow_{4,i,j}\uparrow_{5,i,j}\rangle.
\label{fri}
\end{eqnarray}
This ground state of purely classical nature occurs in the intermediate magnetic-field range delimited by two conditions $h<h_ \mathrm{FRI-PM}$ and $h>h_\mathrm{MD-FRI}$. The FRI phase (\ref{fri}) gives rise to a 3/5-plateau in the zero-temperature magnetization curve,  in accordance with the Lieb-Mattis theorem~\cite{lie62}.

In the magnetic-field range delimited by the conditions $h<h_\mathrm{MD-FRI}$ and $h<h_\mathrm{MD-PM}$, one instead encounters the MD ground state with fully polarized Ising spins and the Heisenberg spins forming dimer singlets, 
\begin{eqnarray}
|{\rm MD} \rangle = \prod_{i,j=1}^L \! |\!\!\uparrow_{1,i,j}\rangle &\!\otimes& \!\frac{1}{\sqrt{2}}(|\!\!\uparrow_{2,i,j}\downarrow_{3,i,j}\rangle 
- |\!\!\downarrow_{2,i,j}\uparrow_{3,i,j}\rangle) \nonumber \\
&\!\otimes& \!\frac{1}{\sqrt{2}}(|\!\!\uparrow_{4,i,j}\downarrow_{5,i,j}\rangle - |\!\!\downarrow_{4,i,j}\uparrow_{5,i,j}\rangle), \nonumber \\
\label{md}
\end{eqnarray} 
which gives rise to a 1/5-plateau in the zero-temperature magnetization curve.
The projection of zero effective field (\ref{zef}) on the $J_2/J_1-h/J_1$ plane, denoted by the blue dashed line in Fig.~\ref{gspd}, closely follows the zero-temperature phase boundary between the FRI and MD ground-state regimes.

We note that the ground-state phase diagram of the Ising-Heisenberg model, shown in  Fig.~\ref{gspd},  is remarkably similar to that of the Heisenberg model
on the diamond-decorated square lattice, see Fig.~3 of Ref.~\onlinecite{cac22}. In particular, both models feature an extended first-order quantum phase transition line that separates a ferrimagnetic regime from the MD regime (in the Heisenberg case, the ferrimagnetic phase still exhibits quantum fluctuations~\cite{cac22}, in contrast to the purely classical nature of the FRI ground state). As we will show in the following, we  observe  similar interesting thermal physics emerging from this first-order line, which we can access by exact methods in the case of the  Ising-Heisenberg model. This adds a very valuable understanding of the underlying thermal physics in both models. To complete the comparison of the two models, we note 
that both models also exhibit a high-field PM phase, whereas the narrow additional dimer-tetramer and spin-canted phases that appear in the Heisenberg case are absent in the Ising-Heisenberg case, due to the pure longitudinal interaction of the Ising spins. Therefore, an analysis of the  Ising-Heisenberg case, as provided here, is of separate value in view of the general relevance of such models for actual compounds, as mentioned above. 

\begin{figure}[t!]
\begin{center}
\includegraphics[width=\columnwidth]{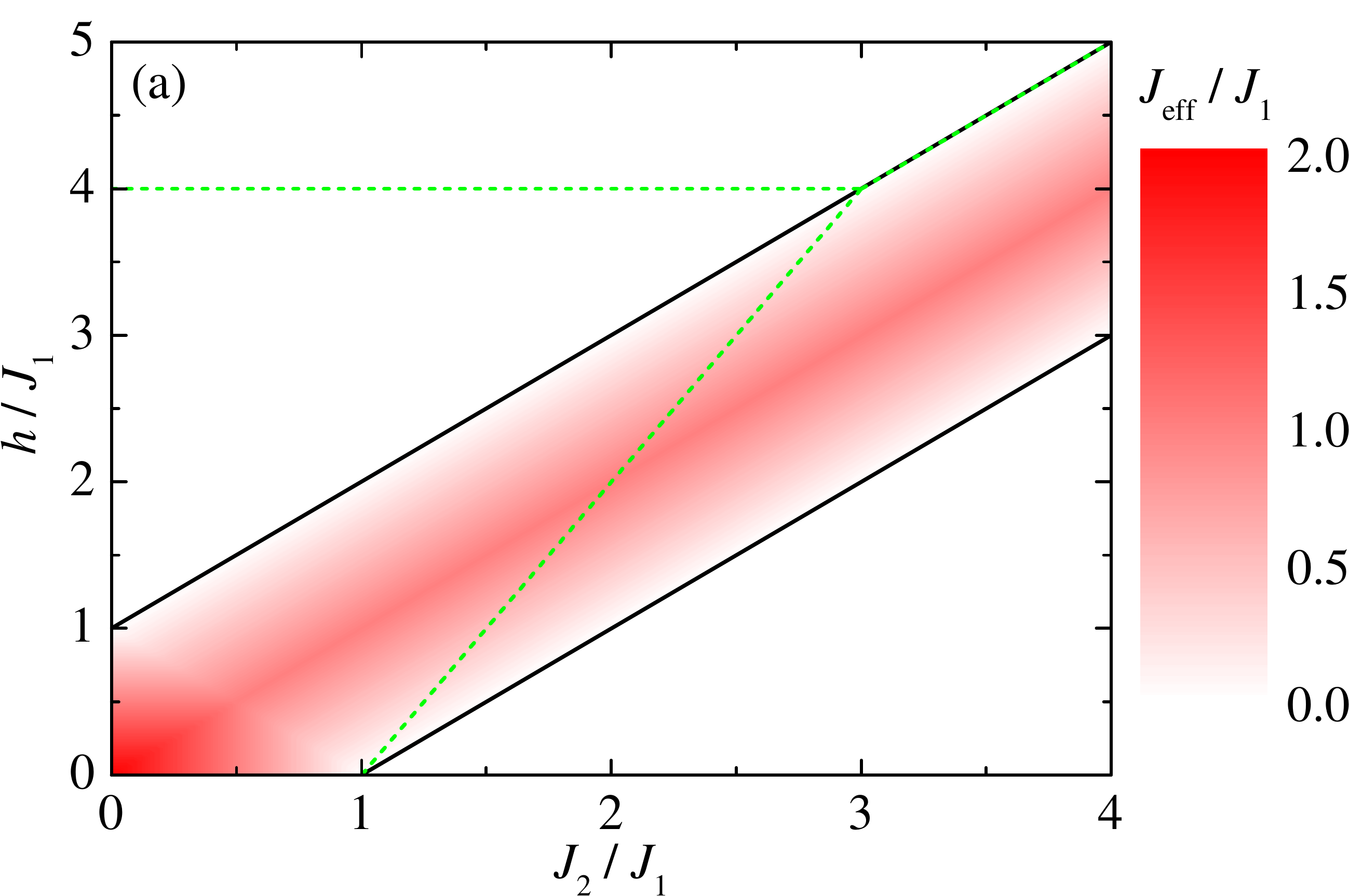}
\includegraphics[width=\columnwidth]{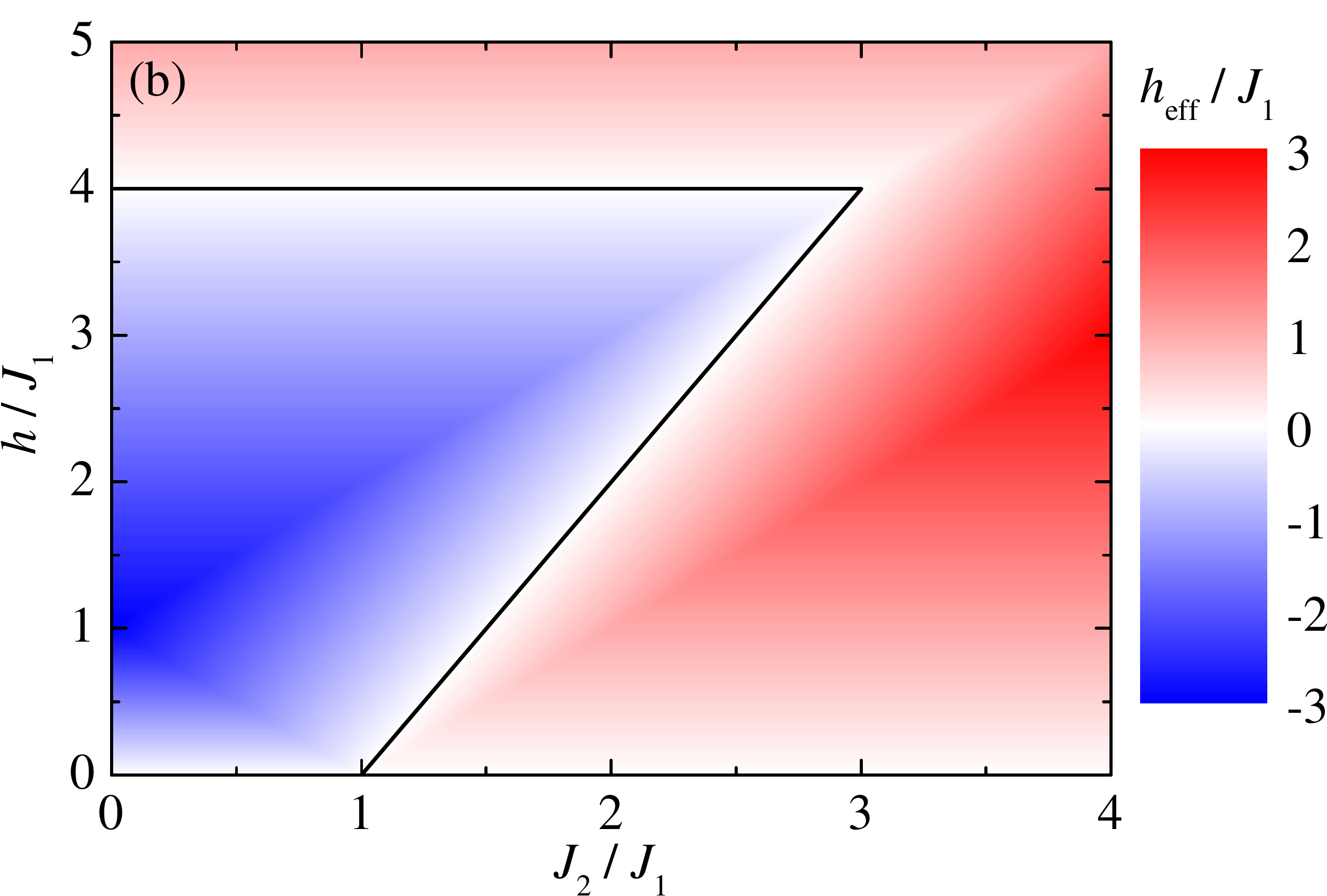}
\end{center}
\vspace{-0.5cm}
\caption{Zero-temperature asymptotic values of the effective interaction $J_{\rm eff}$ (a) and the effective magnetic field $h_{\rm eff}$ (b) in the $J_2/J_1-h/J_1$ plane. Black solid lines determine zero contour lines for the effective interaction $J_{\rm eff} = 0$ and the effective field $h_{\rm eff} = 0$. Green dotted lines in (a) denote the ground-state phase boundaries.}
\label{figef}       
\end{figure}

\subsection{Zero-temperature limit of effective parameters}
Figure \ref{figef} shows the zero-temperature asymptotic values of the effective interaction $J_{\rm eff}$ and the effective field $h_{\rm eff}$, according to the exact mapping  (\ref{zm}) between the spin-1/2 Ising-Heisenberg model on the decorated square lattice and the spin-1/2 Ising model on the square lattice. The presence of  spontaneous long-range order of the latter effective model 
requires zero effective field $h_{\rm{eff}}=0$ in combination with nonzero effective interaction $J_{\rm{eff}}\neq 0$ \cite{ons44}. It is 
clear from Fig.\ \ref{figef}(a) that the regime of nonzero effective interaction $J_{\rm{eff}}\neq 0$ is limited to a rather narrow strip in the parameter region bounded by the conditions $h > J_2 - J_1$ and $h< J_1 + J_2$. The ground-state phase boundary between the FRI and MD phases is thus the only one along which the requirement of a nonzero effective interaction $J_{\rm{eff}}\neq 0$ is fulfilled. 

The  plot of the effective field $h_{\rm eff}$ displayed in Fig.\ \ref{figef}(b) reveals that the effective field changes  sign across two zero-temperature phase boundaries of the FRI ground state. It actually turns out that the effective field is negative ($h_{\rm eff}<0$) within the FRI phase, whereas it becomes positive ($h_{\rm eff}>0$) within the MD and PM phases. Hence, it follows that two zero contour lines of the effective field $h_{\rm eff}=0$ coincide with the ground-state phase boundaries for the FRI-PM and FRI-MD transitions. 
Recall  that the effective interaction vanishes, $J_{\rm{eff}} = 0$, along the ground-state phase boundary between the FRI and PM phases and consequently, this zero-temperature phase transition  cannot appear at any finite temperature. Contrary to this, the other zero contour line of the effective field $h_{\rm eff}=0$, pertinent to the ground-state boundary between the FRI and MD phases, is accompanied by a nonzero effective interaction $J_{\rm{eff}}\neq 0$.  Hence, an eventual extension of this zero-temperature phase transition to finite temperatures is indeed feasible. 
  
\begin{figure}[t!]
\begin{center}
\includegraphics[width=0.9\columnwidth]{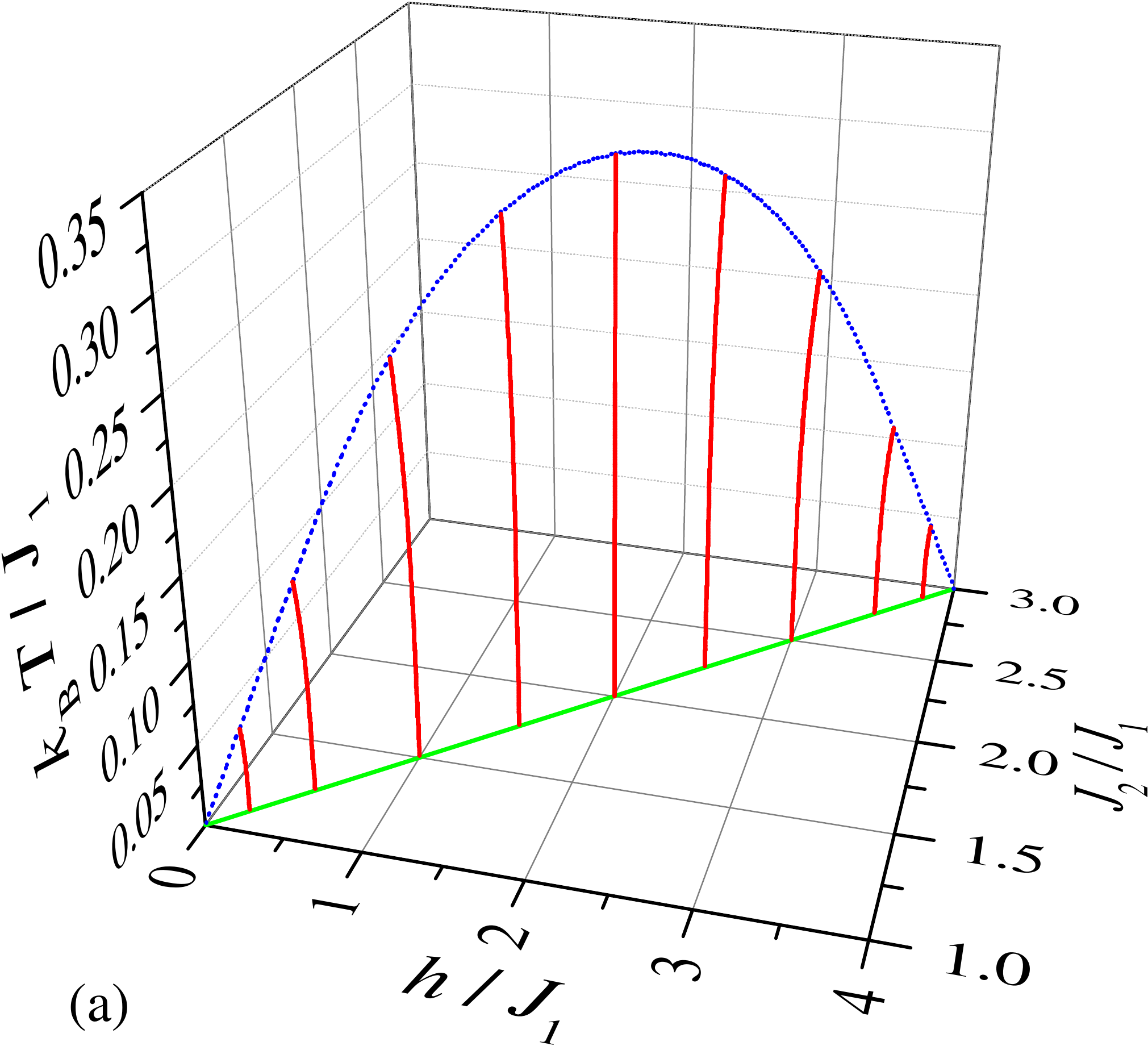} \\[2mm]
\includegraphics[width=0.9\columnwidth]{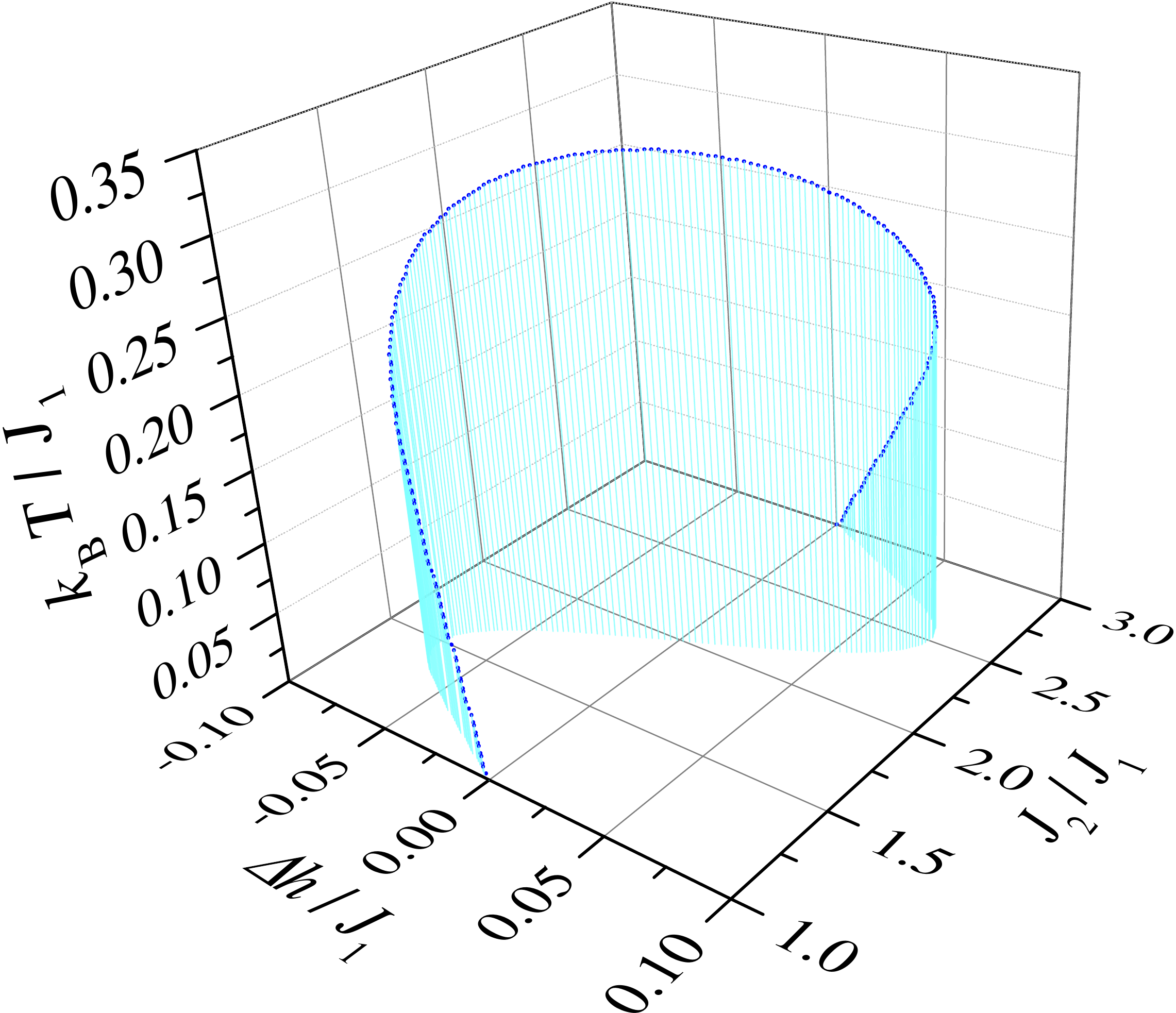}
\end{center}
\vspace{-0.3cm}
\caption{(a) Global phase diagram of the spin-1/2 Ising-Heisenberg model on the diamond-decorated square lattice in the $h/J_1 - J_2/J_1 - k_{\rm B} T/J_1$ parameter space. The blue dotted curve denotes the line of Ising critical points, at which the lines of first-order transitions (red solid lines) between the FRI and MD phases terminate. The green solid line in the $h/J_1 - J_2/J_1$ plane determines the FRI/MD ground-state phase boundary; (b) The blue dotted curve shows 
the line of Ising critical points as a function of the magnetic-field change $\Delta h/J_1$ and the interaction ratio $J_2/J_1$, whereby vertical lines are respective projections towards the $\Delta h/J_1 - J_2/J_1$ plane. The magnetic-field change $\Delta h/J_1$ is given by the difference of the magnetic-field coordinate of the Ising critical point and the respective FRI/MD ground-state phase boundary.}
\label{3dpd}       
\end{figure}

\subsection{Exact results for thermal phase transitions}
Next, we  examine in detail the thermal phase diagram of the spin-1/2 Ising-Heisenberg model on the diamond-decorated square lattice, which is depicted in Fig.\ \ref{3dpd}(a) in the  three-dimensional parameter space $h/J_1-J_2/J_1-k_{\rm B}T/J_1$. In agreement with the ground-state phase diagram shown in Fig.\ \ref{gspd}, the green solid line that falls in Fig.\ \ref{3dpd}(a) onto the $h/J_1-J_2/J_1$ plane corresponds to zero-temperature first-order quantum phase transitions between the FRI and MD phases. It should be pointed out, however, that the coexistence of the FRI and MD phases manifested through a zero effective field, $h_{\rm eff}=0$, is not 
confined to zero temperature, because the nontrivial solution of the zero-field condition (\ref{zef}) extends to finite temperatures as well. It actually turns out that the effective field vanishes $h_{\rm eff}=0$ along an extended surface, i.e.,  a wall of discontinuities. This wall spreads across the red solid curves that are selectively shown in
Fig.\ \ref{3dpd}(a), each representing  a line of thermal first-order phase transitions between the FRI and MD phases for a  selected value of the interaction ratio $J_2/J_1$. These displayed lines of thermal discontinuous phase transitions terminate in a line of Ising critical points unambiguously determined by the critical condition (\ref{critical}),  forming  a line of continuous (second-order) phase transitions, displayed in Fig.\ \ref{3dpd}(a) as a blue dotted line. 

The lines of discontinuous phase transitions (i.e., the red solid curves) are not perfectly vertical, which means that the magnetic field ascribed to the coexistence of the FRI and MD phases  slightly bends upon increasing the temperature while the interaction ratio $J_2/J_1$ is kept constant. To illustrate this more explicitly, Fig.\ \ref{3dpd}(b) shows a three-dimensional plot of the line of the Ising critical points (\ref{critical}) together with its zero-temperature projection onto the $J_2/J_1-\Delta h/J_1$ plane. Here, $\Delta h/J_1$ is given by the difference of the magnetic-field value of the Ising critical point (\ref{critical}) and the zero-temperature phase transition between the FRI and MD phases for a given value of the interaction ratio.
Figure \ref{3dpd}(b) shows that this magnetic-field shift is negative, $\Delta h/J_1<0$, for sufficiently small values of the interaction ratio $J_2/J_1 < 2$, while it becomes positive in the reverse case $J_2/J_1 > 2$. This change implies a  temperature-induced bending of the lines of first-order transitions to lower (higher) magnetic fields for sufficiently small (high) values of the interaction ratio $J_2/J_1 < 2$ ($J_2/J_1 > 2$). Note that similar bending behavior of discontinuous phase transitions was previously reported for several frustrated spin-1/2 Heisenberg models, based on extensive numerical calculations \cite{jim21,sta18,web22,web22b,cac22}, and we will therefore address this intriguing feature in more detail based on  our exact results for the model at hand. 

\begin{figure*}[t!]
\begin{center}
\includegraphics[width=\columnwidth]{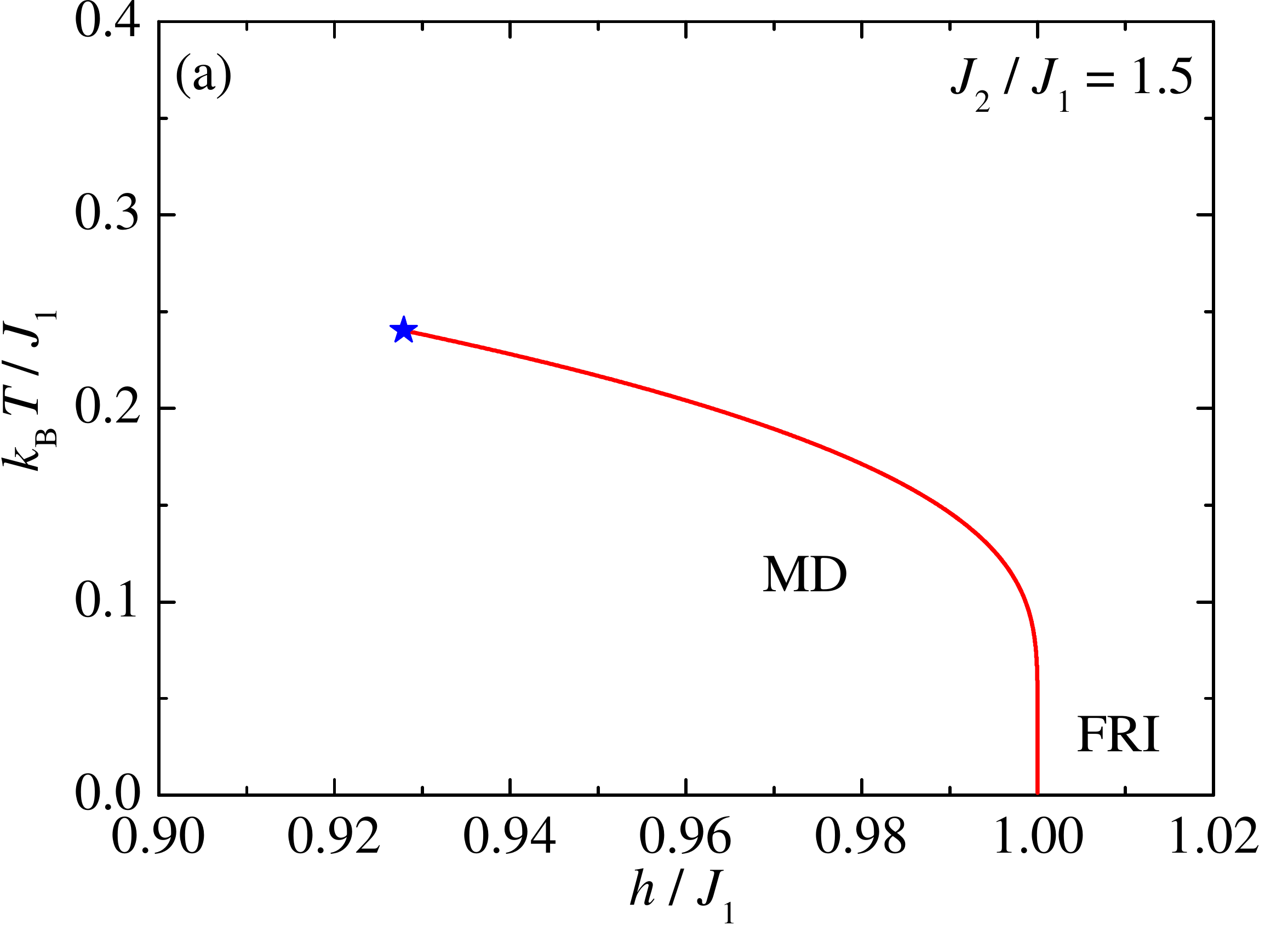}\hfill%
\includegraphics[width=\columnwidth]{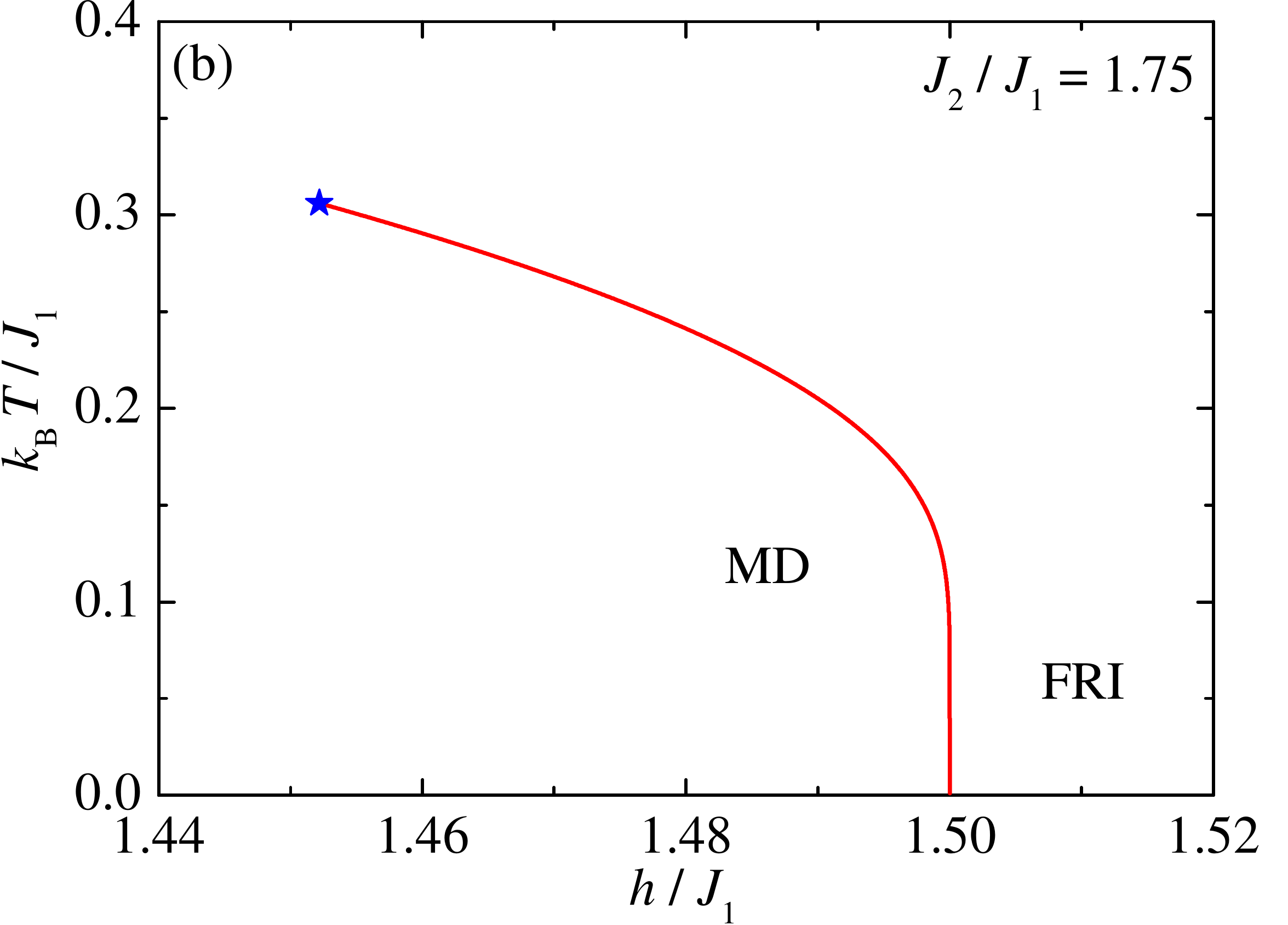}
\includegraphics[width=\columnwidth]{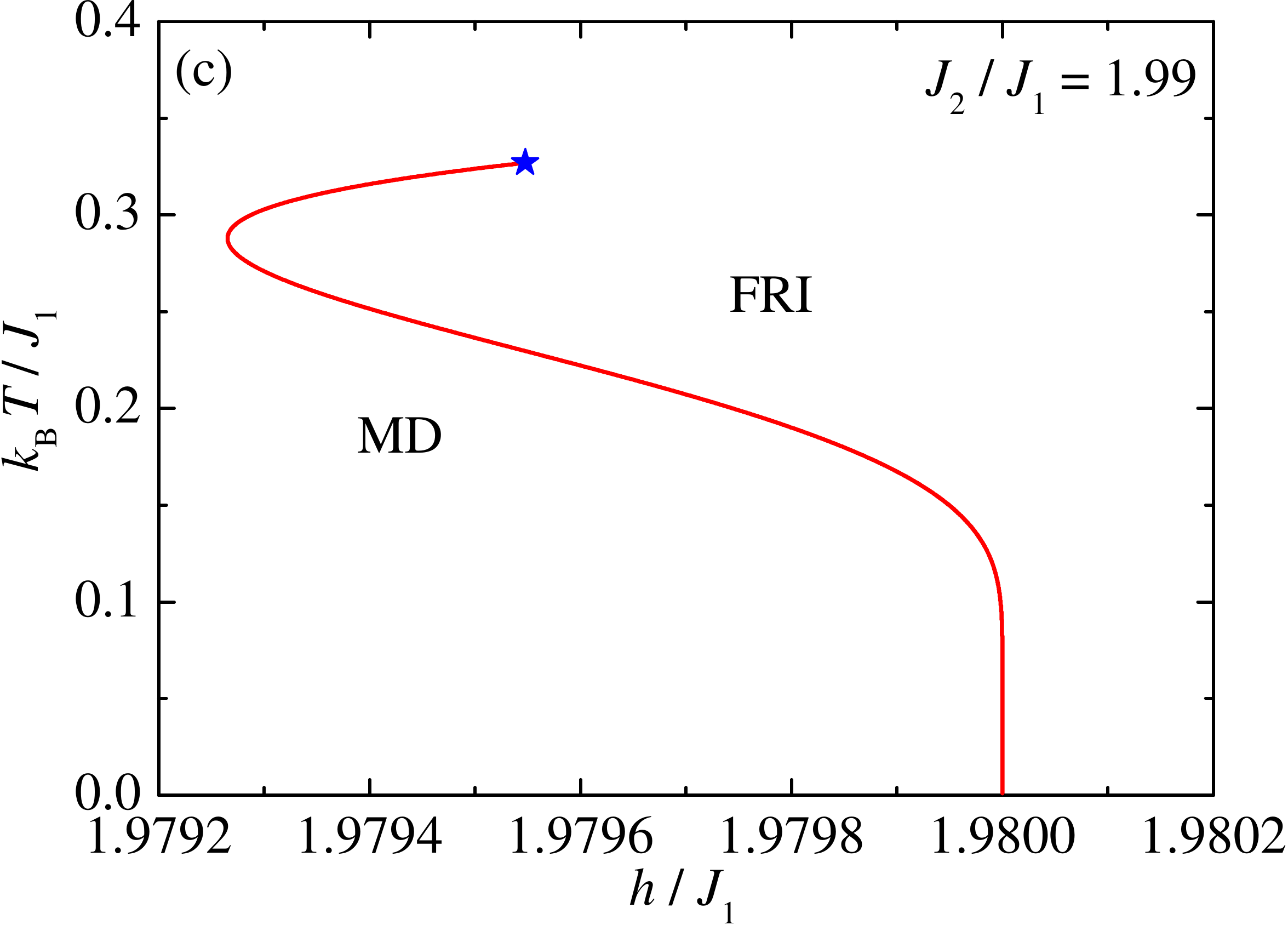}\hfill%
\includegraphics[width=\columnwidth]{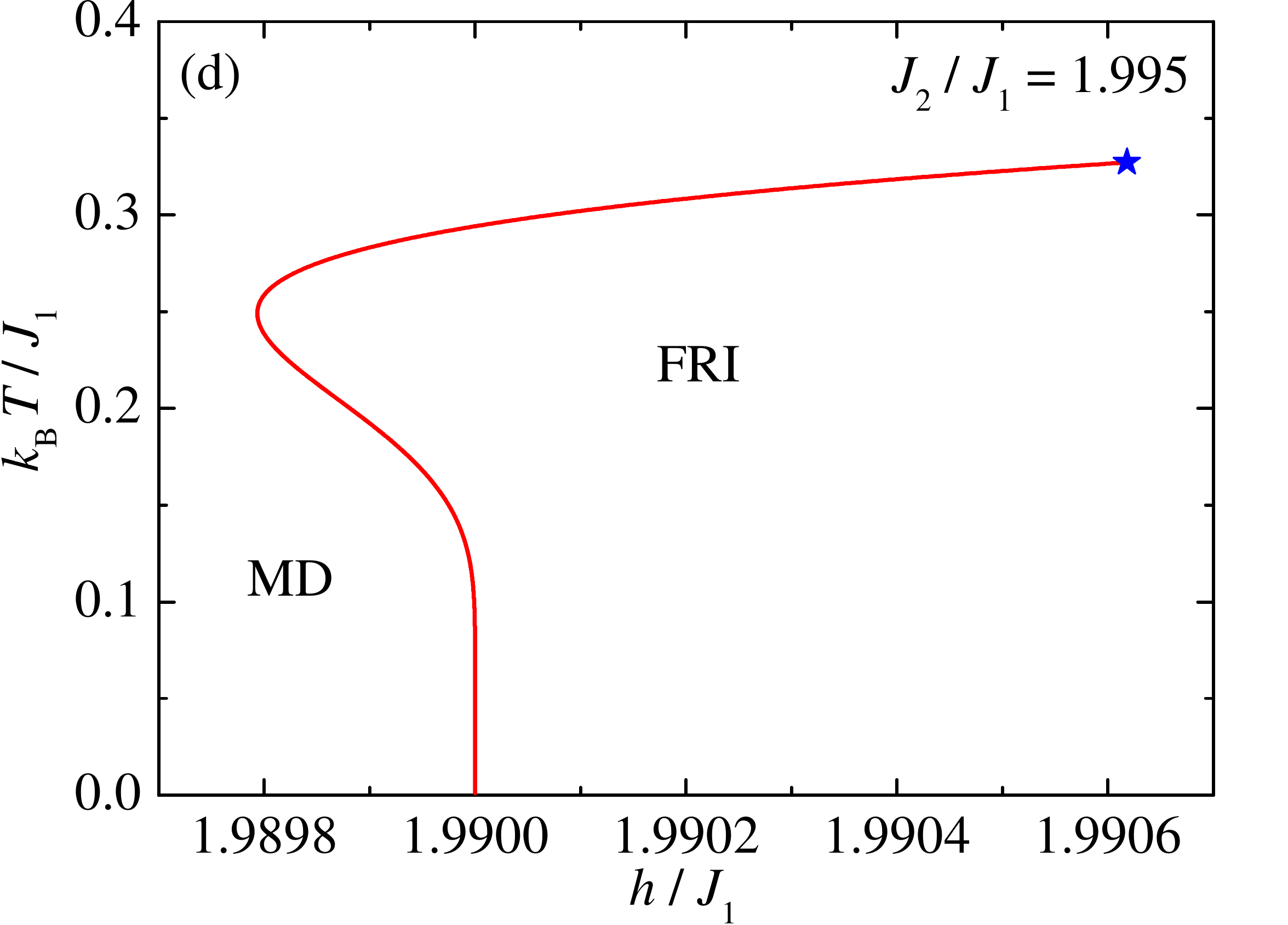}
\includegraphics[width=\columnwidth]{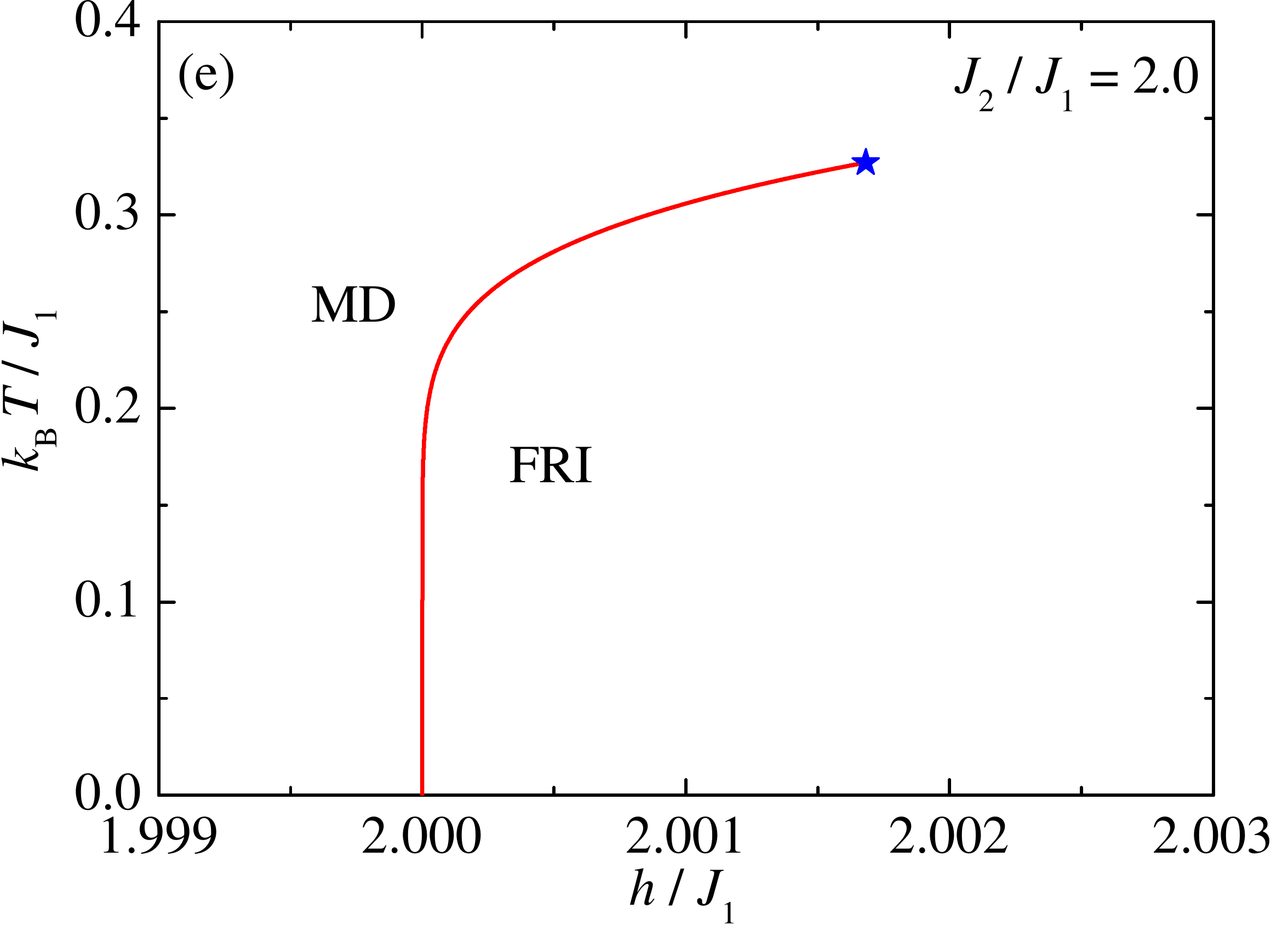}\hfill%
\includegraphics[width=\columnwidth]{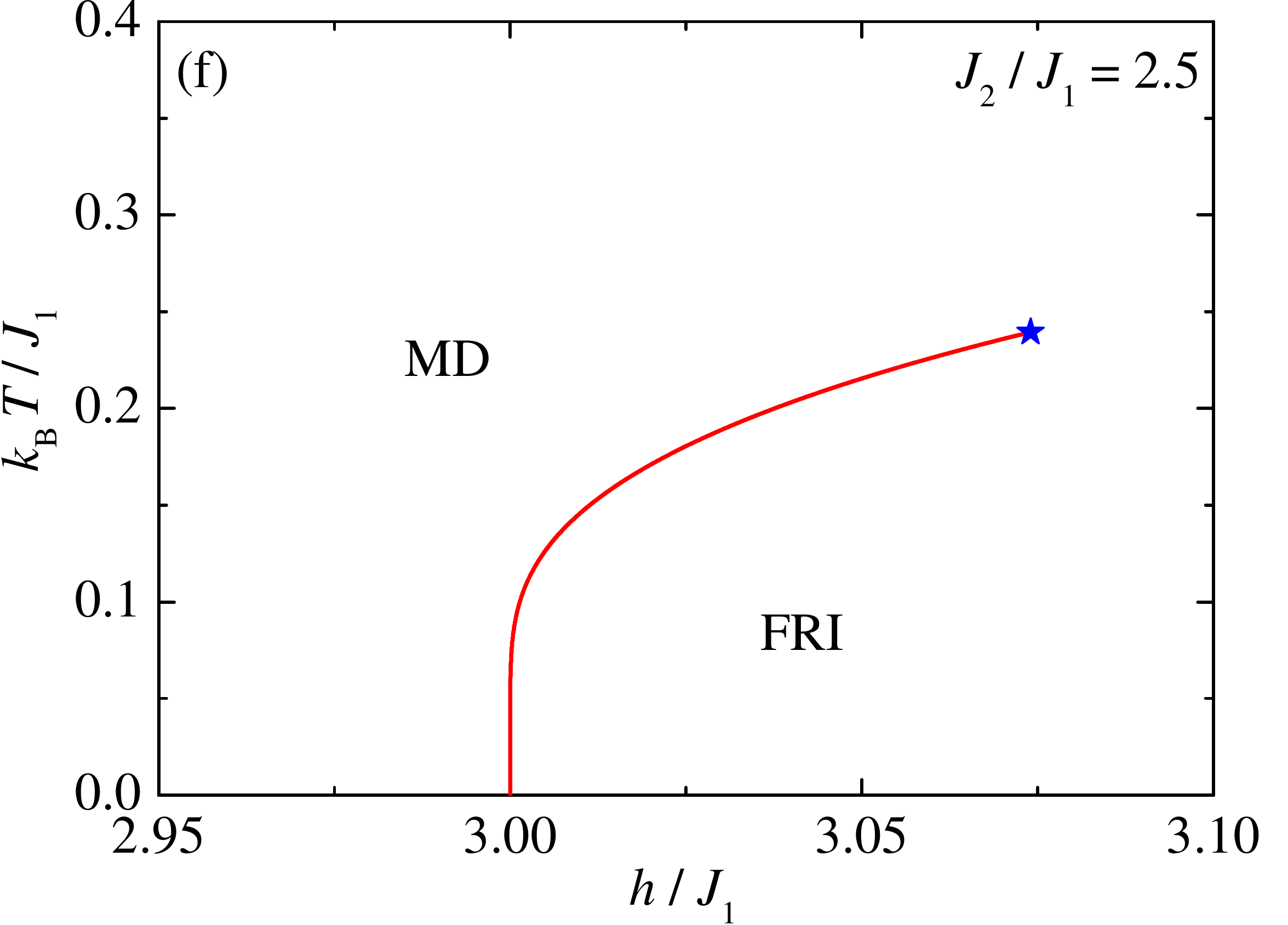}
\end{center}
\vspace{-0.5cm}
\caption{Exact finite-temperature phase diagrams of the spin-1/2 Ising-Heisenberg model on the diamond-decorated square lattice in the magnetic field versus temperature plane at six selected values of the interaction ratio: (a) $J_2/J_1 = 1.5$, (b) $J_2/J_1 = 1.75$, (c) $J_2/J_1 = 1.99$, (d) $J_2/J_1 = 1.995$, (e) $J_2/J_1 = 2.0$, (f) $J_2/J_1 = 2.5$. The solid red curves determine the lines of thermal first-order transitions between the FRI and MD phases, which end at the Ising critical point,  indicated  by a blue star.}
\label{ftpd}       
\end{figure*}

To provide a more comprehensive picture, the lines of the thermal first-order transitions between the FRI and MD phases are plotted in Fig.\ \ref{ftpd} along with the respective Ising critical point for six selected values of the interaction ratio $J_2/J_1$. It directly follows from Fig.\ \ref{ftpd} that the spin-1/2 Ising-Heisenberg model on the diamond-decorated square lattice  exhibits a rather rich diversity of finite-temperature phase diagrams, depending on the coupling ratio $J_2/J_1$: For $J_2/J_1 < 2$ one observes, at low temperatures, a bending of the phase-transition line towards lower magnetic fields, i.e.,  thermal fluctuations prefer the FRI phase over the MD phase [see Figs.\ \ref{ftpd}(a)--(d)]. On the other hand, the line of first-order transitions bends towards higher magnetic fields for larger values of the interaction ratio $J_2/J_1 \geq 2$, giving rise to  a thermally-assisted proliferation of the MD phase at the expense of the FRI phase [see Figs.\ \ref{ftpd}(e) and (f)] instead. 

The most remarkable phase boundary can be found for an interaction ratio $J_2/J_1 \lesssim 2$, as illustrated by two particular cases shown in Figs.\ \ref{ftpd}(c) and (d). Although one may still detect a bending of the phase-transition line towards lower magnetic fields at sufficiently low temperatures, the relevant line of discontinuous phase transitions starts to bend  at higher temperatures in the opposite direction, i.e., towards higher magnetic fields. As a consequence one  surprisingly finds, in a relatively narrow range of magnetic field strength, two consecutive reentrant thermal phase transitions from the MD phase to the FRI phase and vice versa. For the particular choice of the interaction ratio $J_2/J_1 = 1.995$ and a magnetic field of $h/J_1=1.9899$ [Fig.\ \ref{ftpd}(d)], the MD phase is for instance preserved up to the first transition temperature $k_{\rm B}T/J_1\approx 0.1875$. The FRI phase is consecutively realized at moderate temperatures ranging up to the second transition temperature $k_{\rm B}T/J_1\approx 0.285$, and finally the MD phase is  recovered at higher temperatures. Although numerous exact studies have previously reported continuous reentrant phase transitions in two-dimensional exactly solved Ising spin models (see Refs.~\onlinecite{lie86,die20} and references cited therein) since the early discovery of this peculiar phenomenon by Vaks, Larkin, and Ovchinnikov,\cite{vak66} the exact determination of discontinuous reentrant phase transitions 
has, to the best of our knowledge, not been reported 
so far.

\subsection{Bending of the first-order transition lines}
Next, we aim to clarify 
why the discontinuous phase-transition lines between the FRI and MD phases  bend either towards lower or higher magnetic fields. At low  temperature,  the form of the  phase-transition lines is a direct consequence of a few low-energy excitations. To provide a more in-depth understanding, it is convenient to start from a diagonal form of the local Hamiltonians (\ref{eq:hamdsc}) of the diamond spin clusters:
\begin{eqnarray}
\label{ham_diamond}
e_{i,j}^h &=& J_1 S_{23, i, j}^z (S_{1,i,j}^z + S_{1,i+1,j}^z) + \frac{J_2}{2} S_{23, i, j}(S_{23, i, j}+1) \nonumber\\
          &&- \frac{3J_2}{4} - h S_{23,i,j}^z - \frac{h}{4} (S_{1,i,j}^z + S_{1,i+1,j}^z), \nonumber\\
e_{i,j}^v &=& J_1 S_{45, i, j}^z (S_{1,i,j}^z + S_{1,i,j+1}^z) + \frac{J_2}{2} S_{45, i, j}(S_{45, i, j}+1) \nonumber\\
          &&- \frac{3J_2}{4} - h S_{45,i,j}^z - \frac{h}{4} (S_{1,i,j}^z + S_{1,i,j+1}^z), 
\end{eqnarray}
which are expressed in terms of the quantum spin numbers of ${\mathbf S}_{23,i,j} = {\mathbf S}_{2,i,j} + {\mathbf S}_{3,i,j}$ and ${\mathbf S}_{45,i,j} = {\mathbf S}_{4,i,j} + {\mathbf S}_{5,i,j}$ determining the total spin of the dimers and their $z$-components. The overall energy can then be
obtained as a sum of all energy contributions of separate diamond spin clusters (\ref{ham_diamond}).

The ground-state phase diagram shown in Fig.\ \ref{gspd}  involves three different phases: 
\begin{itemize}
	\item The MD phase ($S_{23,i,j}=S_{23,i,j}^z=0$, $S_{1,i,j}^z=1/2$) with the energy per diamond spin cluster of \\
	$$e_\mathrm{MD} = -\frac{3}{4}J_2-\frac{h}{4}.$$ 
	\item The FRI phase ($S_{23,i,j}=S_{23,i,j}^z=1$, $S_{1,i,j}^z=-1/2$) with the energy per diamond spin cluster of 
	$$e_\mathrm{FRI}=\frac{J_2}{4} - J_1 - \frac{3h}{4}.$$
	\item The PM phase ($S_{23,i,j}=S_{23,i,j}^z=1$, $S_{1,i,j}^z=1/2$) with the energy per diamond spin cluster of
	$$e_\mathrm{PM} = \frac{J_2}{4} + J_1 - \frac{5h}{4}.$$
\end{itemize}

Here, we follow the procedure given in the Supplemental Material of Ref.~\onlinecite{sta18}. The low-temperature thermodynamic properties are determined solely by the low-energy excitations above the ground state, such that the free energy $F$ can  be approximated as:
\begin{eqnarray}
\label{fen_app}
F \approx E_{0} - k_{\rm B} T\sum_{m} N_{m}e^{-\beta \Delta_m},
\end{eqnarray}
where $E_0$ denotes the ground-state energy and $\Delta_m$ and $N_m$ denote the energy and the degeneracy of $m$-th excitation. If one considers the phase boundary between the MD and FRI phases, the line of  first-order phase transitions can be obtained from the equality of the free energies of both phases, i.e., $F_\mathrm{MD} = F_\mathrm{FRI}$, like in the case of the Heisenberg model \cite{cac22}.

\begin{figure}
\includegraphics[width=\columnwidth]{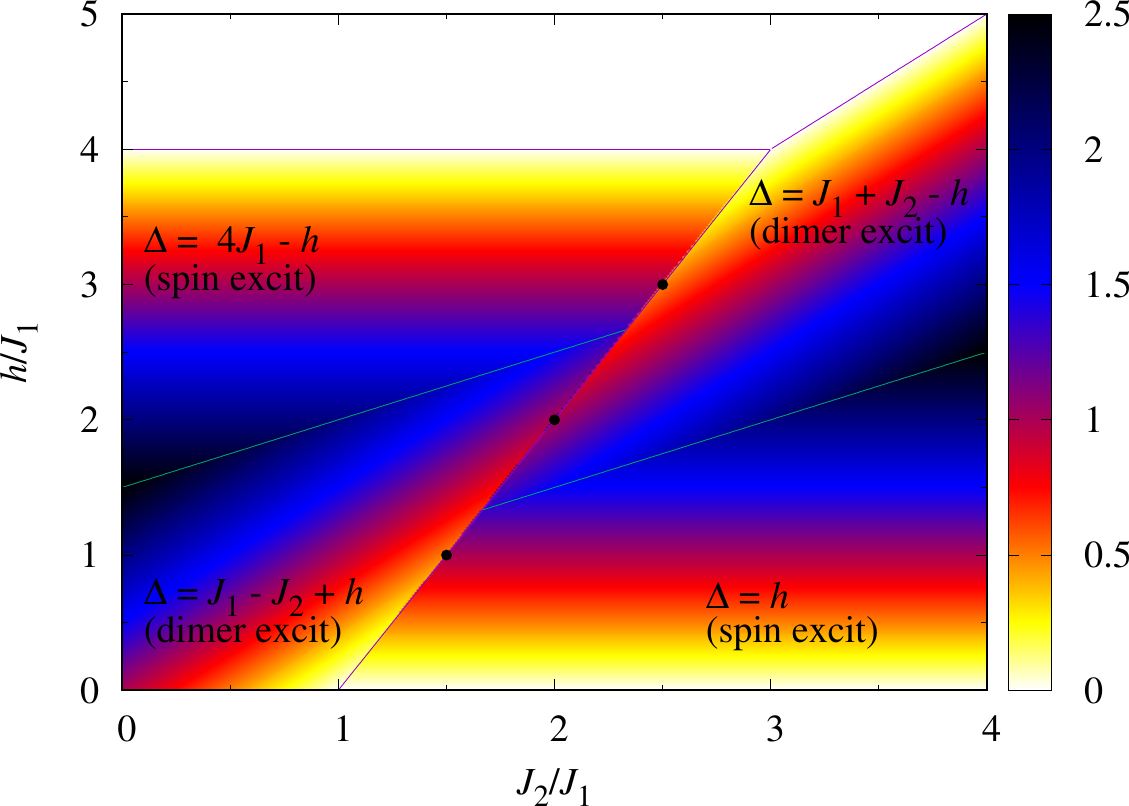}
\caption{Density plot of the lowest excitation energy above the MD and FRI ground states.
Here, `excit.' abbreviates excitations.}
\label{gaps}       
\end{figure}

To proceed further, we require the lowest-energy excitations above a given ground state. First, let us introduce a convenient notation for the eigenenergies of the diamond spin clusters: $E(S_{1,i,j}^z,\{S_{23},S_{23}^z\},S_{1,i,j+1}^z)$ and their respective excitation energies $\Delta E(S_{1,i,j}^z,\{S_{23},S_{23}^z\},S_{1,i,j+1}^z)$. If an excited state relates to an elementary excitation of the  Ising spin, the elementary excitation energy should be multiplied by the factor of 4 in order to obtain the overall energy gain, i.e., $\Delta E(S_{1,i,j}^z,\{S_{23},S_{23}^z\},S_{1,i,j+1}^z) = 4(E(S_{1,i,j}^z,\{S_{23},S_{23}^z\},S_{1,i,j+1}^z) - E_{0})$, since each  Ising spin belongs to four different diamond spin clusters. Using Eq.~(\ref{ham_diamond}) one may readily calculate the energies of the low-energy excited states pertinent to the MD phase:
\begin{eqnarray}
\label{MD_excit}
&& E \left(\frac{1}{2},\{0,0\},-\frac{1}{2}\right) = -\frac{3}{4}J_2, \nonumber\\
&& E \left(\frac{1}{2},\{1,1\},\frac{1}{2}\right) = \frac{1}{4}J_2 + J_1 -\frac{5}{4}h, \nonumber\\
&& E \left(\frac{1}{2},\{1,0\},\frac{1}{2}\right) = \frac{1}{4}J_2  -\frac{1}{4}h. 
\end{eqnarray}
Taking into account that $E_{0} = e_\mathrm{MD}$, the corresponding excitation energies relative to the MD phase are then given by the following expressions:
\begin{eqnarray}
\label{MD_excit2}
&& \Delta E \left(\frac{1}{2},\{0,0\},-\frac{1}{2}\right) = \frac{h}{4}\times 4 = h,\nonumber\\
&& \Delta E \left(\frac{1}{2},\{1,1\},\frac{1}{2}\right) = J_1 + J_2 - h,\nonumber\\
&& \Delta E \left(\frac{1}{2},\{1,0\},\frac{1}{2}\right) = J_2. 
\end{eqnarray}
The excitation energy $\Delta E(\frac{1}{2},\{0,0\},-\frac{1}{2})$ ($\Delta E(\frac{1}{2},\{1,1\},\frac{1}{2})$)
is smallest if $h<h_1$ ($h>h_1$) (see Fig.~\ref{gaps}), where 
\begin{eqnarray}
h_1 = \frac{1}{2}(J_1 + J_2). 
\end{eqnarray}
Note that $h_1$ crosses the ground-state phase boundary $h_\mathrm{MD-FRI}$ at $J_2/J_1=5/3$, $h/J_1=4/3$.

According to Eq.~(\ref{ham_diamond}), the energies of the low-energy excited states pertinent to the FRI phase are given by the expressions:
\begin{eqnarray}
\label{FRI_excit}
&& E\left(\frac{1}{2},\{1,1\},-\frac{1}{2}\right) = \frac{1}{4}J_2 - h, \nonumber\\
&& E\left(-\frac{1}{2},\{0,0\},-\frac{1}{2}\right) = -\frac{3}{4}J_2 + \frac{1}{4}h, \nonumber\\
&& E\left(-\frac{1}{2},\{1,0\},-\frac{1}{2}\right) = \frac{1}{4}J_2 + \frac{1}{4}h,
\end{eqnarray}
which are consistent with the excitation energies:
\begin{eqnarray}
\label{FRI_excit2}
&& \Delta E\left(\frac{1}{2},\{1,1\},-\frac{1}{2}\right) = \left(J_1 - \frac{h}{4}\right)\times 4 = 4J_1 - h, \nonumber\\
&& \Delta E\left(-\frac{1}{2},\{0,0\},-\frac{1}{2}\right) = J_1 - J_2 + h, \nonumber\\
&& \Delta E\left(-\frac{1}{2},\{1,0\},-\frac{1}{2}\right) = J_1 + h.
\end{eqnarray}
It follows from Eq.\ (\ref{FRI_excit2}) that the excitation energy $\Delta E(-\frac{1}{2},\{0,0\},-\frac{1}{2})$ ($\Delta E(\frac{1}{2},\{1,1\},-\frac{1}{2})$) is smallest for $h<h_2$ ($h>h_2$) (see Fig.~\ref{gaps}), where
\begin{eqnarray}
h_2 = \frac{1}{2}(3J_1 + J_2).
\end{eqnarray}
Comparing with (\ref{Eq:hMDFRI}) shows that 
$h_2$ crosses the ground-state phase boundary $h_\mathrm{MD-FRI}$ at $J_2/J_1=7/3$, $h/J_1=8/3$.
We note that very similar behavior as the one shown in Fig.~\ref{gaps} for the Ising-Heisenberg model
is found in the Heisenberg model on the diamond-decorated square lattice, see Fig.~11 of Ref.~\onlinecite{cac22}.

Using the excitation energies (\ref{MD_excit2}) and (\ref{FRI_excit2}) one readily finds the low-temperature approximation for the free energy per diamond spin cluster in the MD and FRI phases:
\begin{eqnarray}
f_\mathrm{MD} &=& \frac{F_\mathrm{MD}}{2N} 
\approx e_\mathrm{MD} - \frac{1}{\beta}
\left( \frac{1}{2} {\rm e}^{-\beta\Delta E(\frac{1}{2},\{0,0\},-\frac{1}{2})}
\right.\nonumber\\ && \left.
+ {\rm e}^{-\beta\Delta E(\frac{1}{2},\{1,1\},\frac{1}{2})}
+ {\rm e}^{-\beta\Delta E(\frac{1}{2},\{1,0\},\frac{1}{2})}
\right),
\\
f_\mathrm{FRI} &=& \frac{F_\mathrm{FRI}}{2N} 
\approx e_\mathrm{FRI} - \frac{1}{\beta}
\left( \frac{1}{2} {\rm e}^{-\beta\Delta E(\frac{1}{2},\{1,1\},-\frac{1}{2})}
\right.\nonumber\\ && \left.
+ {\rm e}^{-\beta\Delta E(-\frac{1}{2},\{0,0\},-\frac{1}{2})}
+ {\rm e}^{-\beta\Delta E(-\frac{1}{2},\{1,0\},-\frac{1}{2})}
\right). \qquad
\end{eqnarray}

\begin{figure}[t!]
\includegraphics[width=\columnwidth]{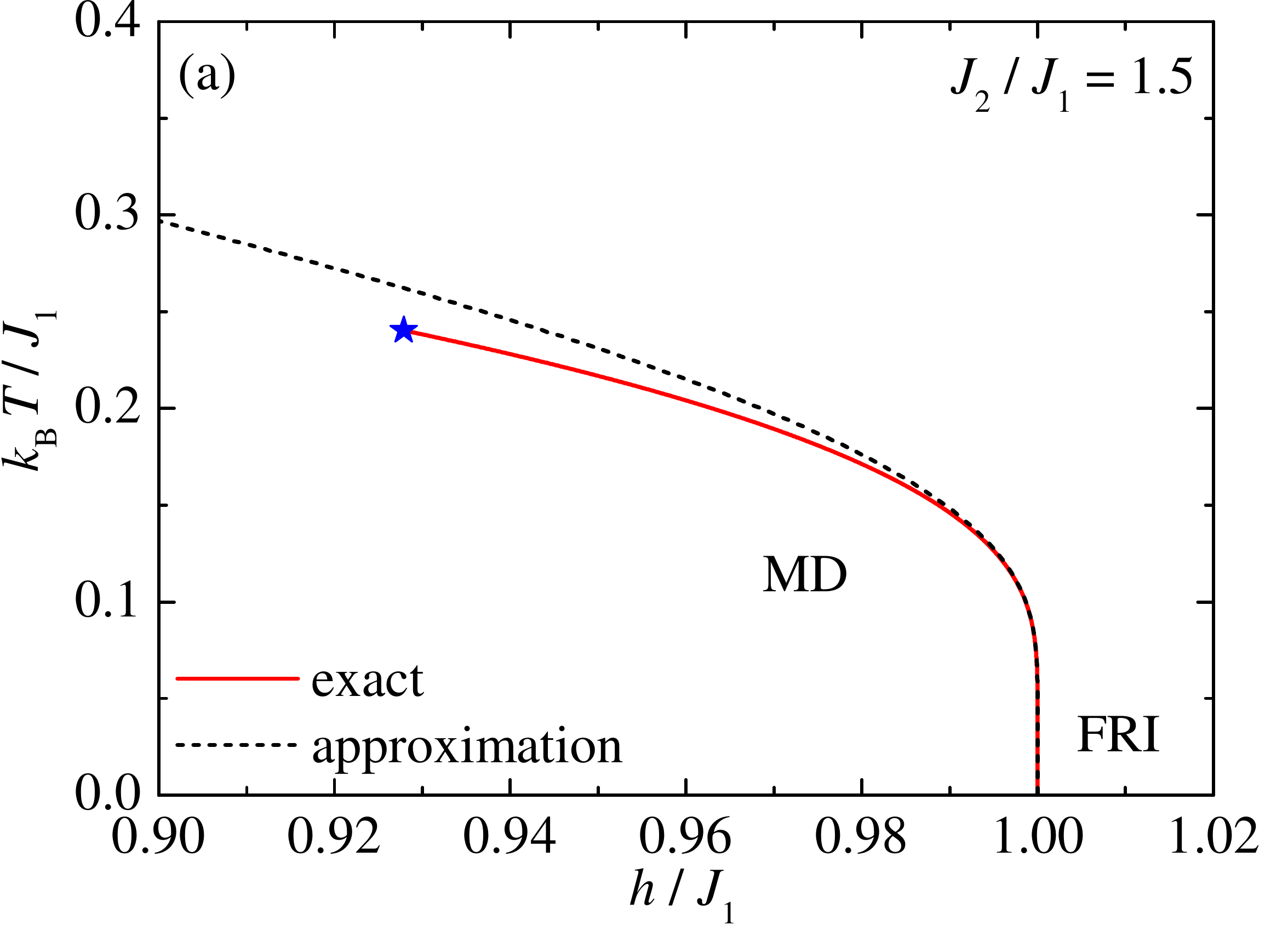}
\includegraphics[width=\columnwidth]{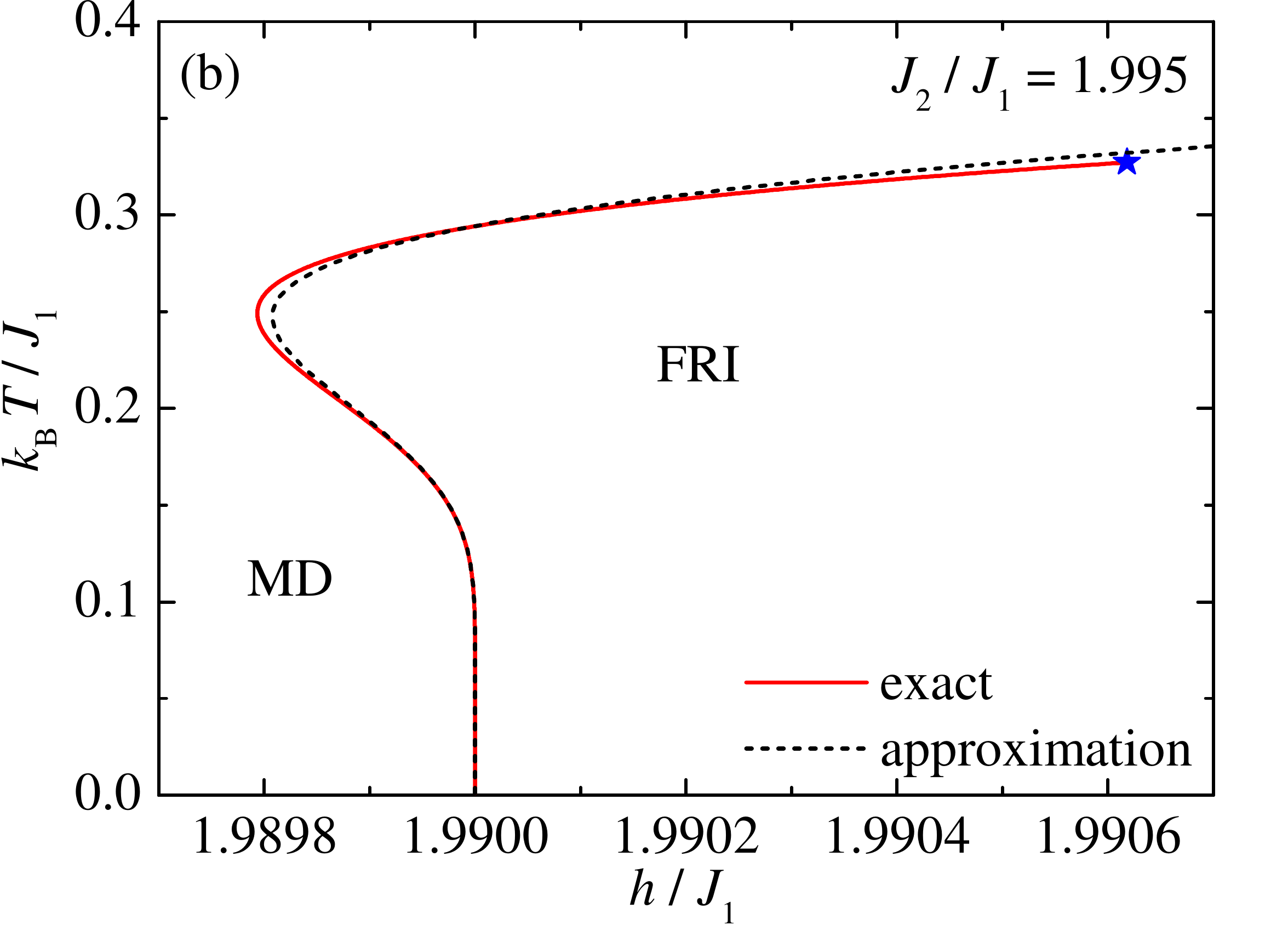}
\includegraphics[width=\columnwidth]{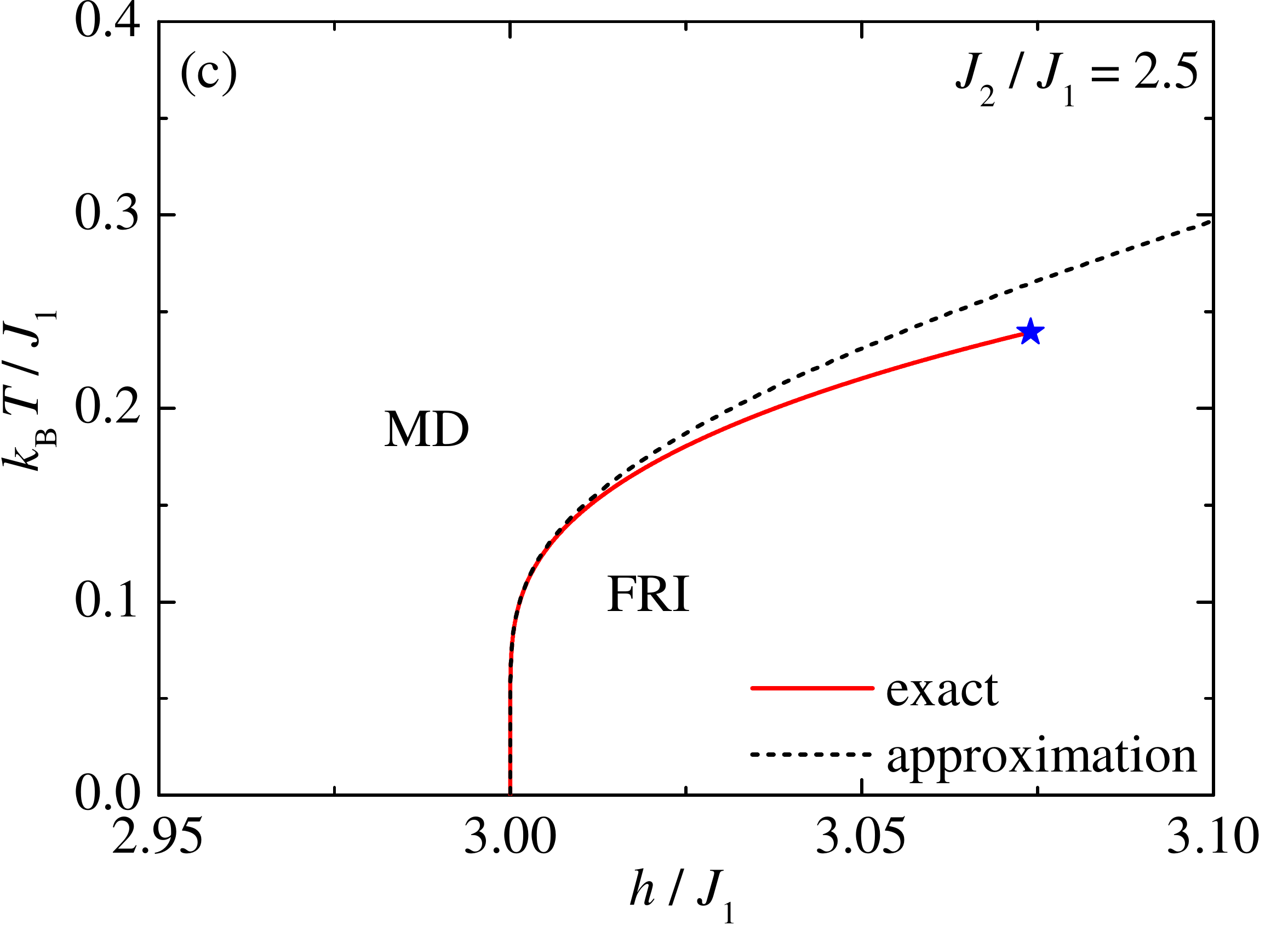}
\caption{Exact results and low-temperature approximation for the thermal first-order phase-transition lines of the spin-1/2 Ising-Heisenberg model on the  diamond-decorated square lattice for three different values of the interaction ratio: (a) $J_2/J_1=1.5$, (b) $J_2/J_1=1.995$, (c) $J_2/J_1=2.5$.}
\label{bend}       
\end{figure}

It is convenient to again introduce the deviation of the magnetic field from the transition field value between the MD and FRI phases,
\begin{eqnarray}
\Delta h = h - h_{\rm MD-FRI} = h - 2(J_2-J_1), 
\end{eqnarray}
which is small at low temperatures. Next, we  solve the equation $f_{\rm MD}=f_{\rm FRI}$ by keeping only  terms linear in $\Delta h$, which gives the following result:
\begin{eqnarray}
\Delta h &\approx & 
2 k_{\rm B} T \left[
{\rm e}^{-\beta(3J_1-J_2)} {+} \frac{1}{2} {\rm e}^{-\beta(2J_2-2J_1)}  {+} {\rm e}^{-\beta J_2}
\right.
\nonumber\\ &&
\left. - {\rm e}^{-\beta(J_2-J_1)} {-} \frac{1}{2}{\rm e}^{-\beta(6J_1-2J_2)} {-} {\rm e}^{-\beta(2J_2-J_1)} 
\right]. \quad \label{Deltah_app}
\end{eqnarray}
This result
contains  contributions from all low-energy excitations given by Eqs.~(\ref{MD_excit2}) and (\ref{FRI_excit2}). However, only a few of these contributions remain relevant for particular values of the magnetic field and the interaction ratio. In Fig.\ \ref{bend}(a) we illustrate the first paradigmatic example, typical for the parameter region $J_2/J_1<5/3$, for which the lowest-energy excitation in the MD phase $\Delta E(\frac{1}{2},\{0,0\},-\frac{1}{2}) = h$ relates to a spin flip of the  Ising spin, whereas the lowest-energy excitation in the FRI phase $\Delta E(-\frac{1}{2},\{0,0\},-\frac{1}{2}) = J_1 - J_2 + h$ correspond to  the Heisenberg dimer (see Fig.~\ref{gaps}). In this respect, one may further simplify Eq.~(\ref{Deltah_app}) by retaining only the two most relevant terms, related to the aforementioned lowest-energy excitations:
\begin{eqnarray}
& \Delta h \approx 2 k_{\rm B} T \! \left[\frac{1}{2} {\rm e}^{-\beta(2J_2-2J_1)} {-}  {\rm e}^{-\beta(J_2-J_1)} \right]\!, 
\frac{J_2}{J_1}{<}\frac{5}{3}.
\label{Deltah_1}
\end{eqnarray}
This equation 
implies that the phase-transition line between the FRI and MD phases  bends to lower magnetic fields, $\Delta h<0$, because the lowest excitation energy in the FRI phase is smaller than the one in the MD phase. This finding is  consistent with the phase-transition lines depicted in Fig.~\ref{bend}(a), whereby the low-temperature approximation (\ref{Deltah_1}) correctly reproduces at sufficiently low temperatures the form of the exact transition line remarkably accurate (note that within this approximation, we cannot determine the position of the critical point). 

For the second paradigmatic example, typical for larger value of the interaction ratio $J_2/J_1>7/3$, the lowest-energy excitation 
$\Delta E \left(\frac{1}{2},\{1,1\},\frac{1}{2}\right) = J_1 + J_2 - h$ of the Heisenberg dimer in the MD phase has a lower energy than the lowest-energy excitation $\Delta E\left(\frac{1}{2},\{1,1\},-\frac{1}{2}\right) = 4J_1 - h$ of the  Ising spins in the FRI phase. If only those two lowest-energy excitations are taken into consideration, the formula (\ref{Deltah_app}) for $\Delta h$ simplifies to 
\begin{eqnarray}
& \Delta h \approx 2 k_{\rm B} T \! \left[ {\rm e}^{-\beta(3J_1-J_2)} {-} \frac{1}{2}{\rm e}^{-\beta(6J_1-2J_2)} \right]\!, \frac{J_2}{J_1}>\frac{7}{3}. \quad
\label{Deltah_2}
\end{eqnarray}
This result shows
that the phase-transition line between the FRI and MD phases now bends to higher magnetic fields $\Delta h>0$, since the lowest excitation energy in the FRI phase is higher than that in the MD phase. This fact explains why the exact phase-transition line displayed in Fig.~\ref{bend}(c) shifts towards higher magnetic fields in accordance with the prediction obtained by the low-temperature approximation (\ref{Deltah_2}).  

\begin{figure*}[t!]
\begin{center}
\includegraphics[width=\columnwidth]{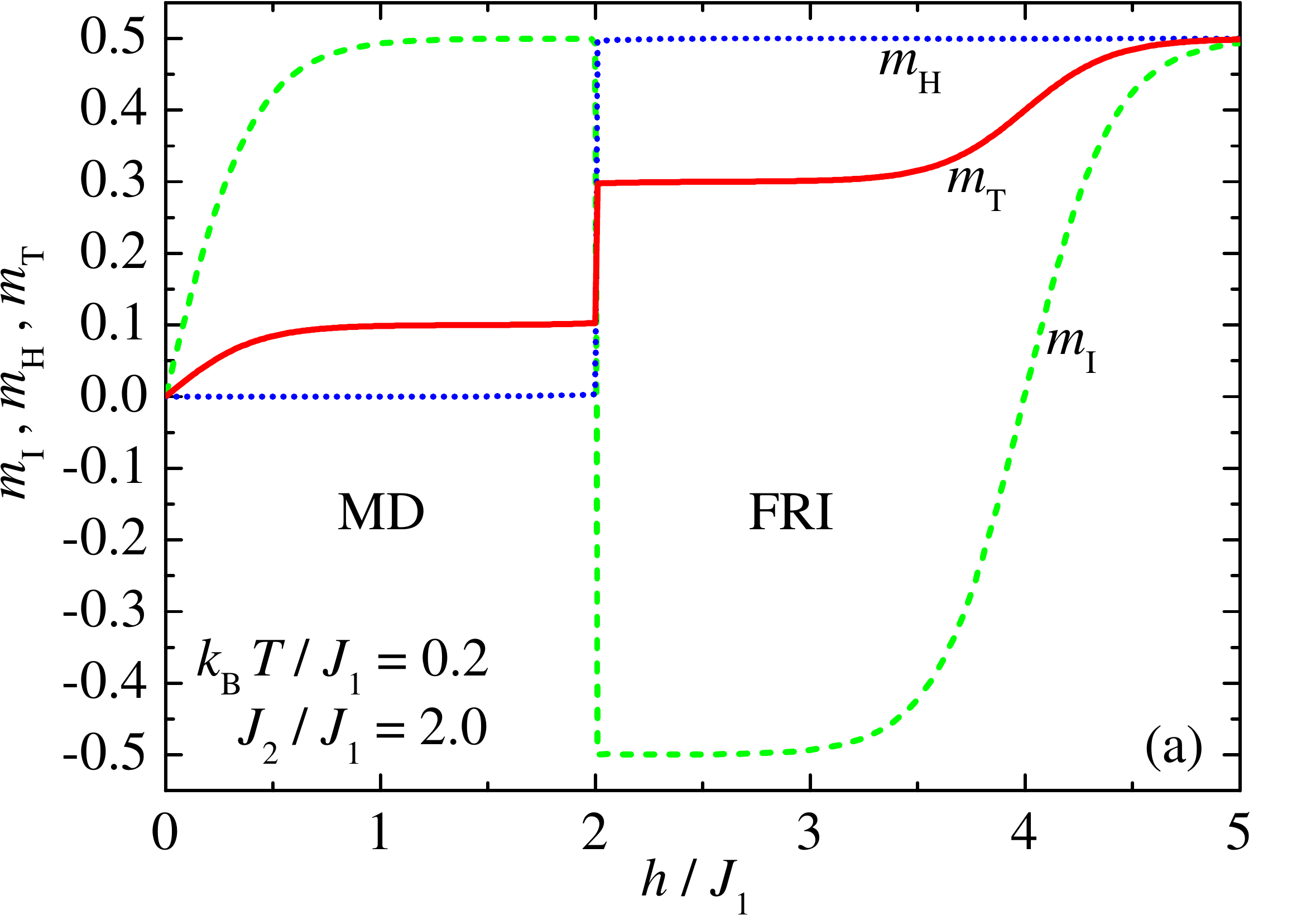}\hfill%
\includegraphics[width=\columnwidth]{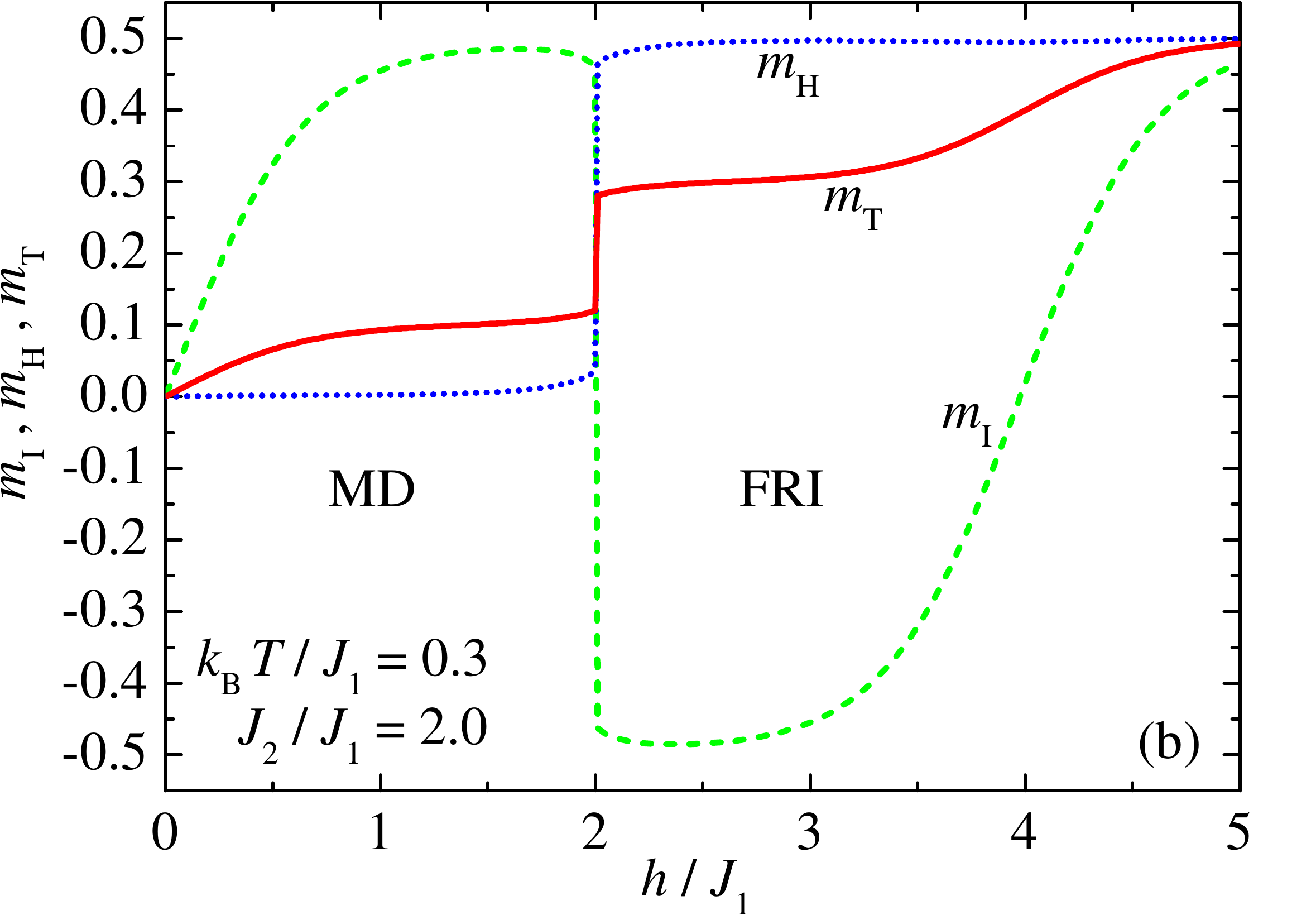} \\[2mm]
\includegraphics[width=\columnwidth]{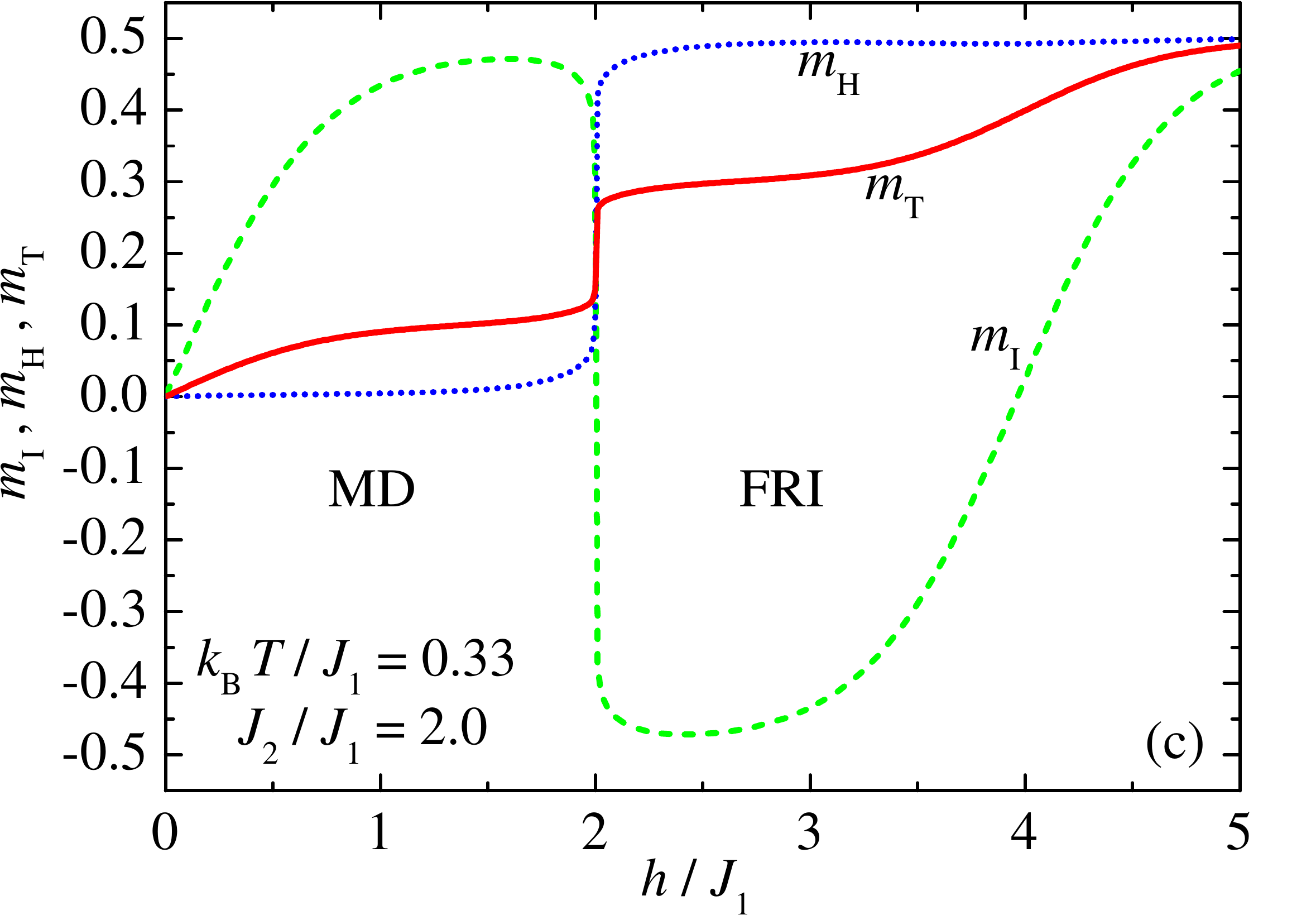}\hfill%
\includegraphics[width=\columnwidth]{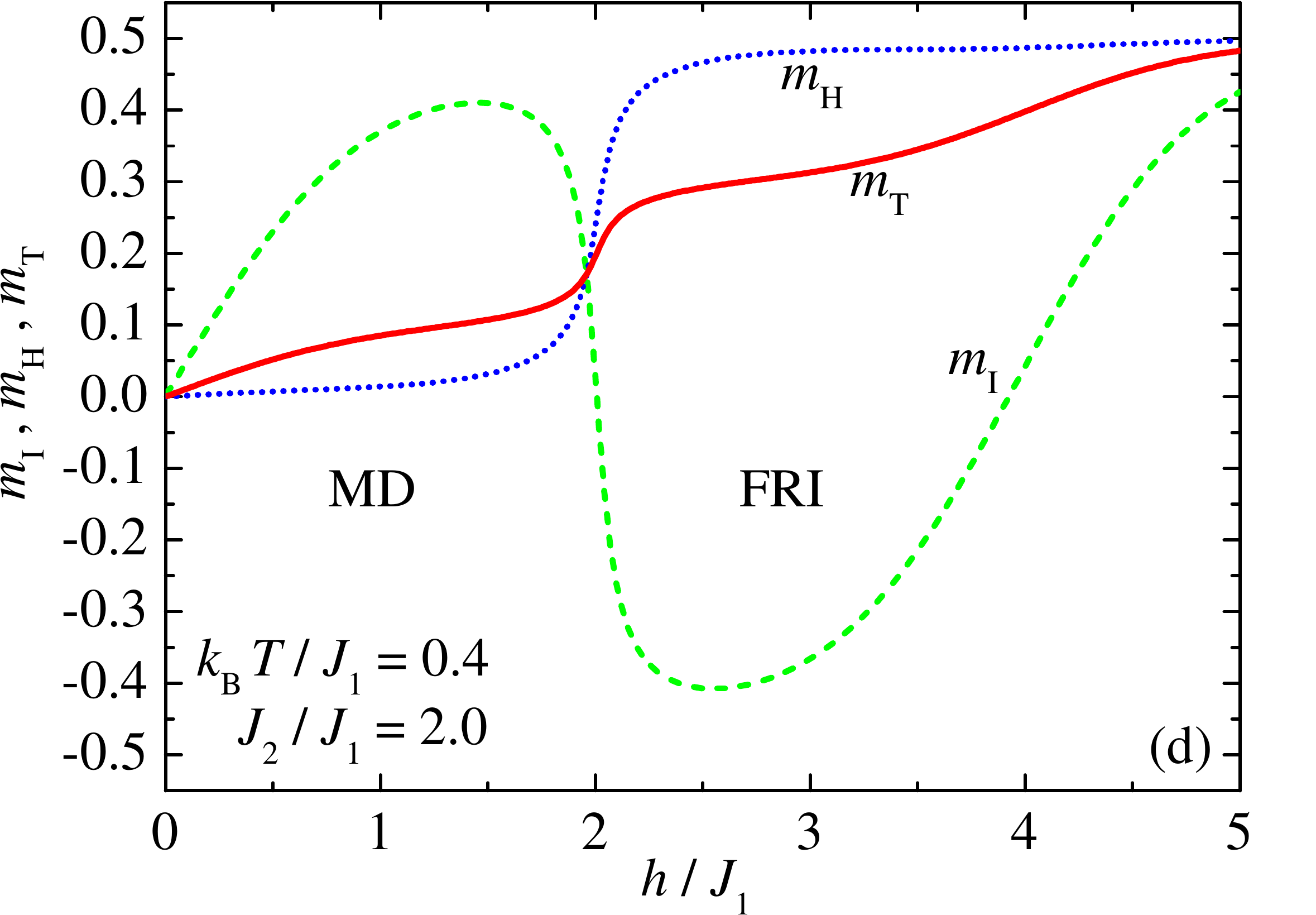}
\end{center}
\vspace{-5mm}
\caption{Results from Monte Carlo simulations for the magnetization curves of the spin-1/2 Ising-Heisenberg model on the diamond-decorated square lattice for a fixed value of the interaction ratio $J_2/J_1 = 2$ and four different values of the temperature: (a) $k_{\rm B}T/J_1 = 0.2$, (b) $k_{\rm B}T/J_1 = 0.3$, (c) $k_{\rm B}T/J_1 = 0.33$, (d) $k_{\rm B}T/J_1 = 0.4$.}
\label{mcmsd}       
\end{figure*}

Similar arguments can also be exploited for moderate values of the interaction ratio $5/3<J_2/J_1<7/3$ excluding the particular case close to $J_2/J_1 = 2$. Under this condition, the lowest-energy excitation in the MD and FRI phases relate to an elementary excitation of the Heisenberg dimer with the relevant excitation energies $\Delta E \left(\frac{1}{2},\{1,1\},\frac{1}{2}\right) = J_1 + J_2 - h$ and $\Delta E(-\frac{1}{2},\{0,0\},-\frac{1}{2}) = J_1 - J_2 + h$, respectively. Thermally-driven changes of the transition field between the MD and FRI phases consequently follow from the formula:
\begin{eqnarray}
&&\Delta h \approx 2 k_{\rm B} T \left[ {\rm e}^{-\beta(3J_1-J_2)} -  {\rm e}^{-\beta(J_2-J_1)} \right],
\nonumber\\
&&\frac{5}{3}<\frac{J_2}{J_1}<\frac{7}{3}, \frac{J_2}{J_1}\neq 2.
\label{Deltah_3}
\end{eqnarray}
which can be either positive or negative, depending on whether the interaction ratio is greater or smaller than $J_2/J_1=2$, respectively. The particular case $J_2/J_1=2$ is 
special, because the first and second lowest-energy excitations in both phases are equal and their contributions in Eq.~(\ref{Deltah_app}) cancel out. Therefore, the third lowest-energy excitation of the Heisenberg dimers to the zero-magnetization triplet state decisively determines the thermal behavior of the magnetic-field change $\Delta h$:
\begin{eqnarray}
& \Delta h \approx 2 k_{\rm B} T \left[ {\rm e}^{-\beta J_2} {-} {\rm e}^{-\beta(2J_2-J_1)} \right], \frac{J_2}{J_1}=2.
\label{Deltah_4}
\end{eqnarray}
Evidently, the lowest excitation energy $\Delta E \left(\frac{1}{2},\{1,0\},\frac{1}{2}\right) = J_2$ in the MD phase is smaller than the one 
$\Delta E\left(-\frac{1}{2},\{1,0\},-\frac{1}{2}\right) = J_1 + h$ in the FRI phase and hence, the phase-transition line between the MD and FRI phases bends towards higher magnetic fields, in accordance with what is observed in Fig.~\ref{ftpd}(e). The specific case with the interaction ratio $J_2/J_1 \lesssim 2$ is even much more intricate because of a subtle interplay between a few low-energy excitations. The phase-transition line between the MD and FRI phases at first bends at low enough temperatures towards lower magnetic fields in accordance with  formula (\ref{Deltah_3}), but an opposite bending towards higher magnetic fields emerges at moderate temperatures, due to a decisive role of the third-lowest excitations of the Heisenberg dimer to the zero-magnetization triplet state. It can thus be concluded that reentrant phase transitions that are observed within a narrow region of the interaction parameters $J_2/J_1 \lesssim 2$ can be attributed to a subtle interplay of all three lowest-energy excitations (\ref{MD_excit2}) and (\ref{FRI_excit2}) above the respective MD and FRI phases [see Fig.~\ref{bend}(b)]. 

\subsection{Monte Carlo simulations}

\begin{figure*}[t!]
\begin{center}
\includegraphics[width=\columnwidth]{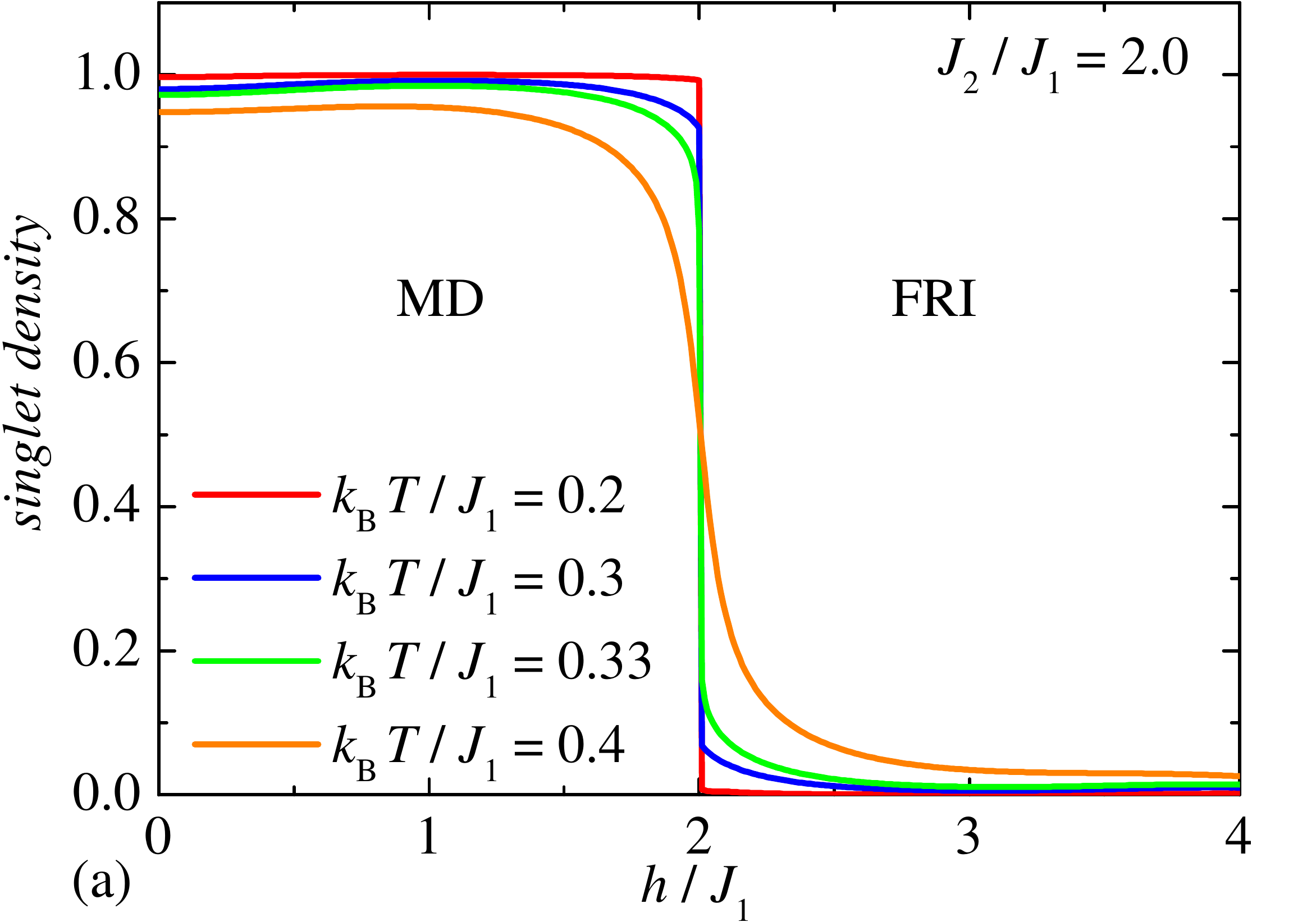}\hfill%
\includegraphics[width=\columnwidth]{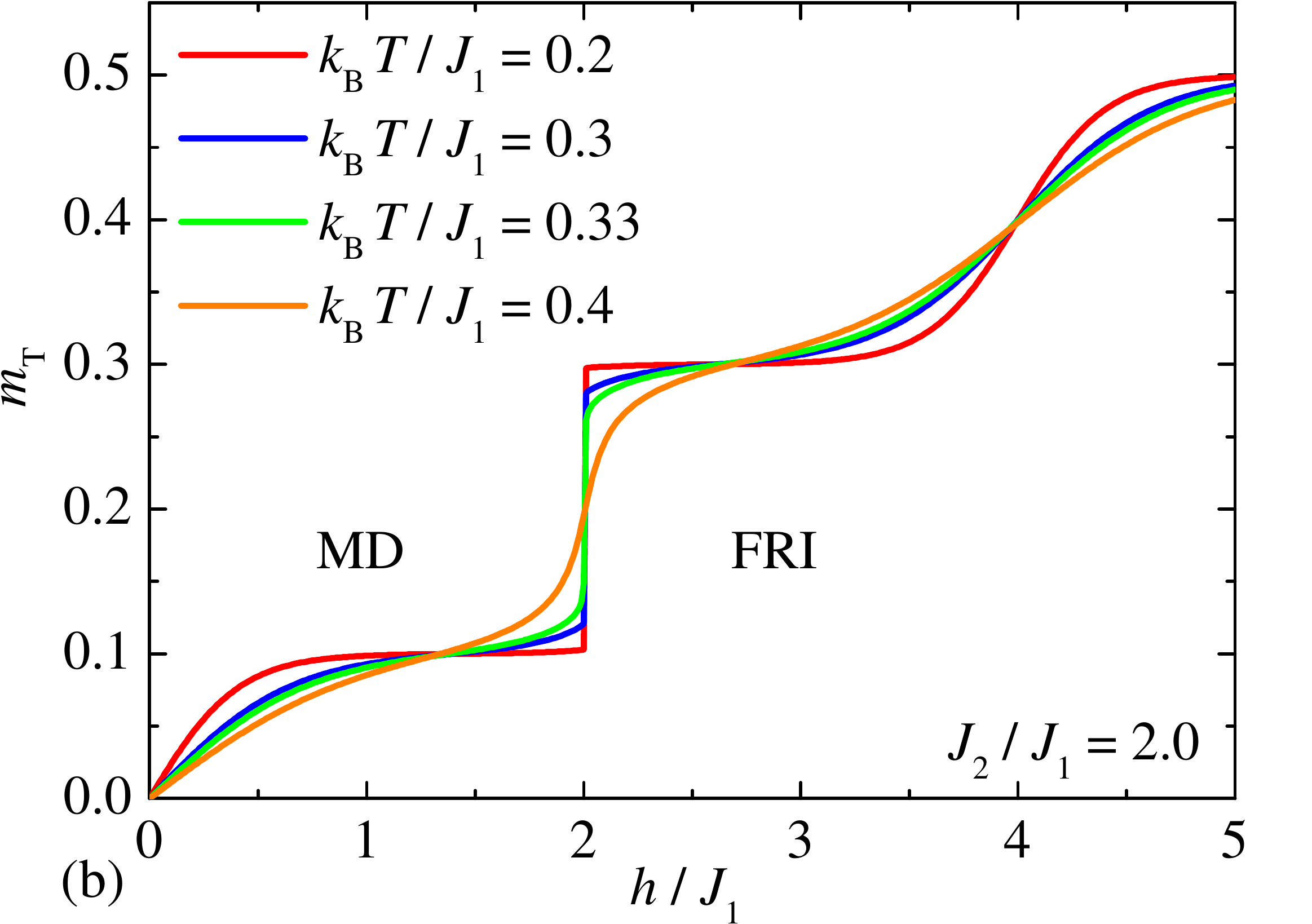}
\end{center}
\vspace{-5mm}
\caption{Magnetic-field dependencies of the singlet density (a) and the total magnetization (b) of the spin-1/2 Ising-Heisenberg model on the diamond-decorated square lattice for a fixed value of the interaction ratio $J_2/J_1=2$ and four different temperatures.}
\label{sdam2}       
\end{figure*}

In order to provide an independent confirmation of the exact results presented in the previous sections, we additionally performed classical Monte Carlo simulations of the effective spin-1/2 Ising model on the square lattice with temperature-dependent nearest-neighbor interaction and magnetic  field, given by the Hamiltonian (\ref{hef}), from which one may easily extract the magnetic behavior of the spin-1/2 Ising-Heisenberg model on the diamond-decorated square lattice for arbitrary temperature and magnetic field. First, our 
attention will be focused on a detailed investigation of magnetization curves, which are depicted in Fig.\ \ref{mcmsd} for a fixed value of the interaction ratio $J_2/J_1=2$ and four different values of temperature. Here, we show the separate magnetizations $m_\mathrm{H}$ and $m_\mathrm{I}$ for the Heisenberg and Ising spins {calculated according to Eqs.\ (\ref{miei}) and (\ref{mh}), respectively,  as well as the total magnetization per spin $m_\mathrm{T} = (m_\mathrm{I} + 4 m_\mathrm{H})/5$.} 

Figure \ref{mcmsd} shows that the magnetization curves of the spin-1/2 Ising-Heisenberg model on the diamond-decorated square lattice exhibit two distinct intermediate 1/5 and 3/5 plateaus whose microscopic nature can be inferred from the local magnetization of the Ising and Heisenberg spins. At sufficiently low magnetic fields $h/J_1<2$, the Ising spins gradually tend towards the magnetic-field direction, while there is almost no contribution from the Heisenberg spins to the total magnetization. These findings conform with the MD phase as being composed of the polarized Ising spins and the singlet-dimer Heisenberg spin pairs. At higher magnetic fields, $h/J_1>2$, the Heisenberg spins are oriented in the magnetic-field direction, in contrast to the Ising spins predominantly pointing in the opposite direction. The observed values of the local magnetization of the Ising and Heisenberg spins thus coincide with the spin arrangement attributed  to the FRI phase. 

In agreement with the exact finite-temperature phase diagram presented in Fig.\ \ref{ftpd}(e), the magnetization curves display a finite magnetization jump at sufficiently low temperatures, related to a discontinuous field-driven phase transition, which gradually shrinks upon increasing the temperature, until it completely disappears at a certain critical temperature pertinent to a continuous field-driven phase transition. The magnetization curves plotted in Fig.\ \ref{mcmsd}(a)--(b) for the  lowest two temperatures actually exhibit quite analogous dependencies with a pronounced magnetization discontinuity, which is subject to a gradual reduction and smoothing upon increasing of temperature. The magnetization curves undergo a crucial change in the close vicinity of the critical temperature $k_{\rm B}T/J_1 \approx 0.33$, at which it exhibits a smooth continuous dependence across the Ising critical point, instead of the discontinuous magnetization jump. The Ising critical point is a prominent manifestation of the continuous field-driven phase transition, which is represented by an inflection point with an infinite tangent in the relevant magnetization curve [see Fig.\ \ref{mcmsd}(c)]. Contrary to this, the magnetization curves recorded at temperatures exceeding the critical temperature are smooth and continuous functions of the magnetic field, without any singular point as exemplified in Fig.\ \ref{mcmsd}(d) for the particular case of $k_{\rm B}T/J_1 = 0.4$. Although there does not appear any magnetic-field driven phase transition, the local and total magnetization still exhibit a crossover  between the values typical for the MD and FRI phases as some residual trace of the low-temperature phase transition.
A similar crossover  can be also found between the FRI and PM phases around the magnetic field $h/J_1 = 4$. However, the crossover between the FRI and PM phases emerges for arbitrarily low temperature, as a consequence of  the absence of any magnetic-field driven phase transition between the FRI and PM phases at any nonzero temperature. 

The MD and FRI phases essentially differ in the character of the spin arrangement of the Heisenberg dimers. {This means that the difference of the densities of singlets in the MD and FRI phases plays the role of an order parameter for the MD/FRI phase transition in close analogy to the difference in density of the vapor and liquid phases of water.}
As a matter of fact, at $T=0$, all Heisenberg spin pairs create singlet-dimer states in the MD phase where the singlet density thus equals to one ($n=1$), while  in the FRI all Heisenberg spins are  in the polarized triplet state, so that the singlet density becomes zero ($n=0$).

\begin{figure*}[t!]
\begin{center}
\includegraphics[width=\columnwidth]{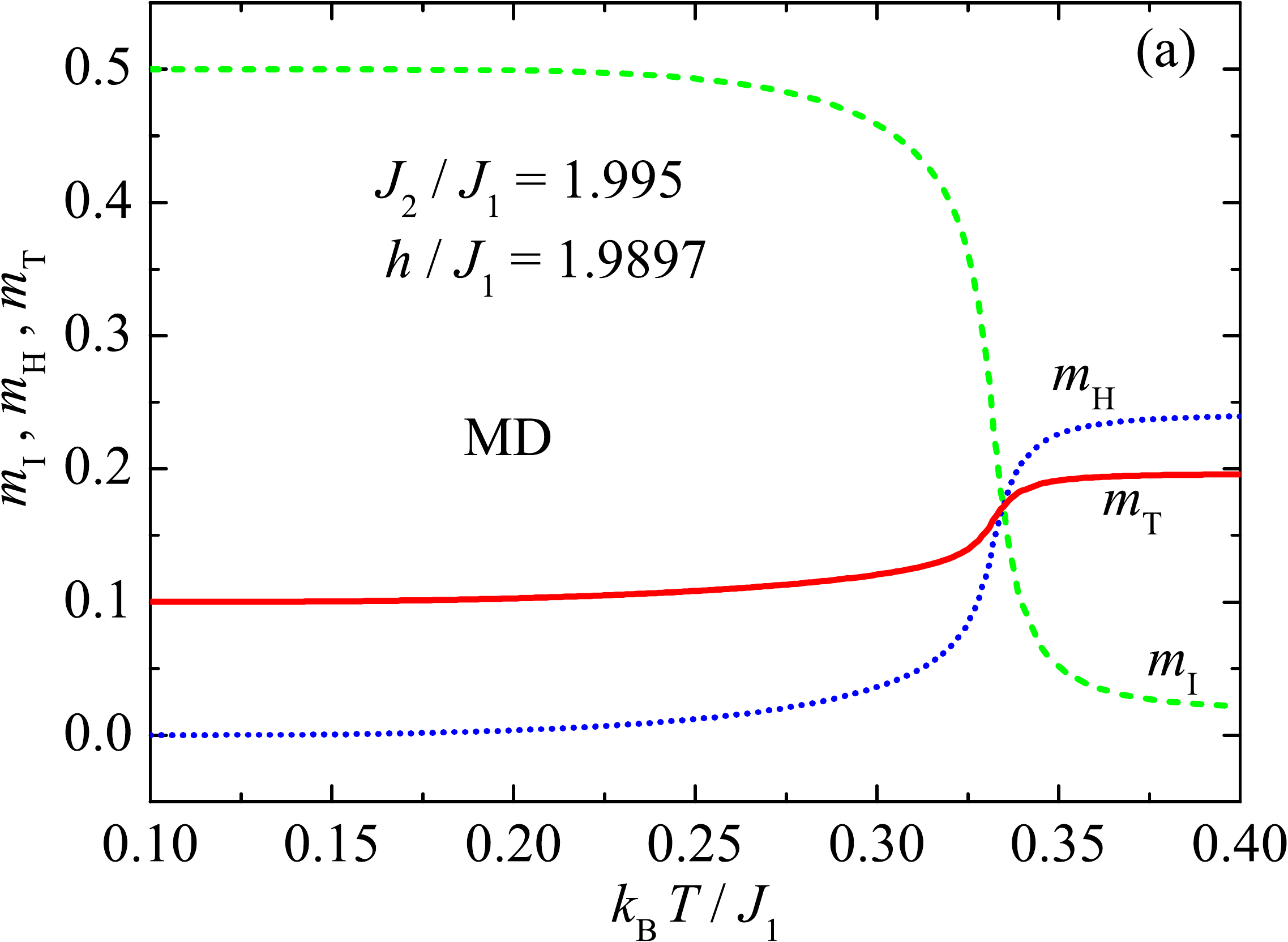}\hfill%
\includegraphics[width=\columnwidth]{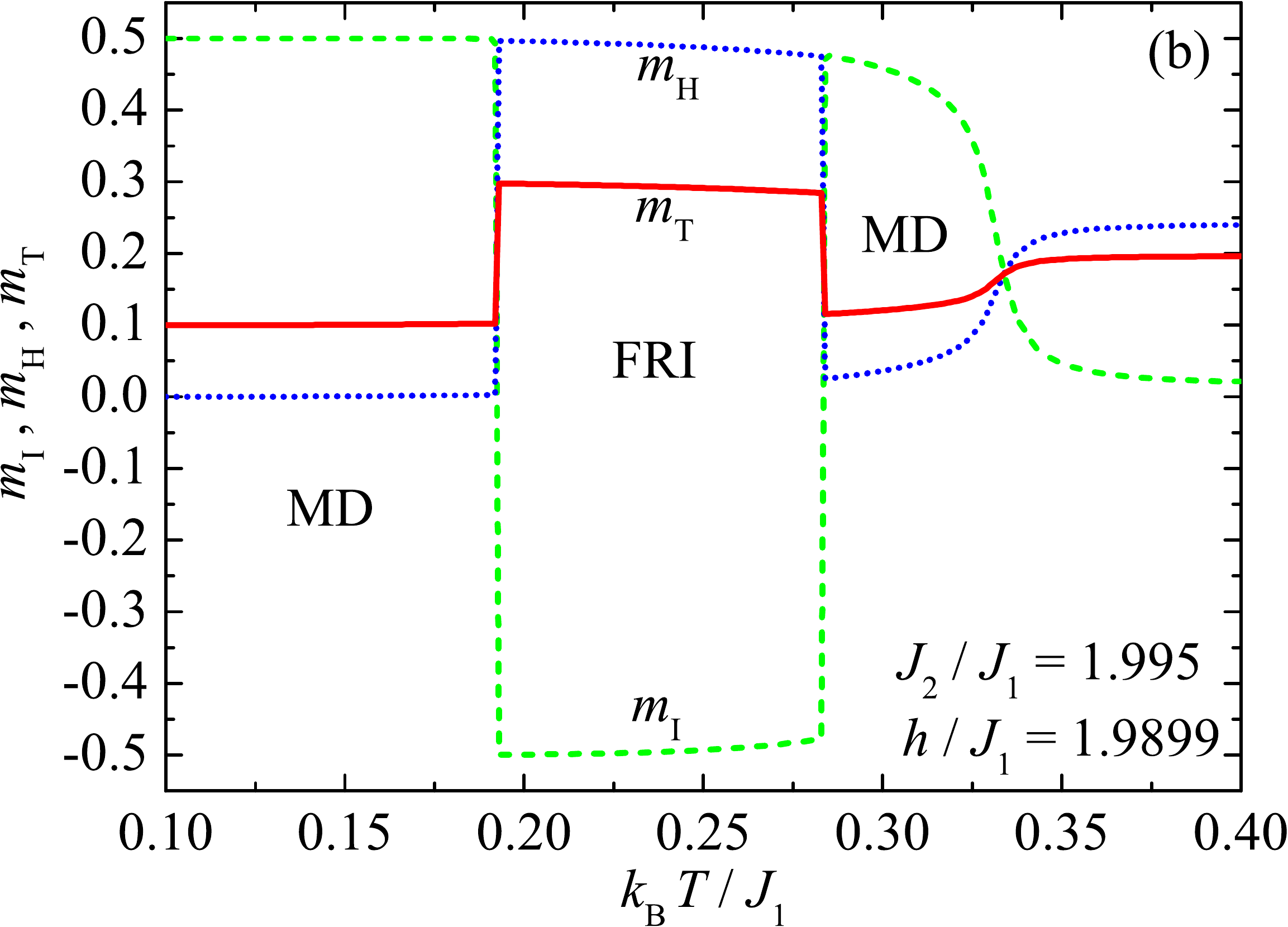} \\[2mm]
\includegraphics[width=\columnwidth]{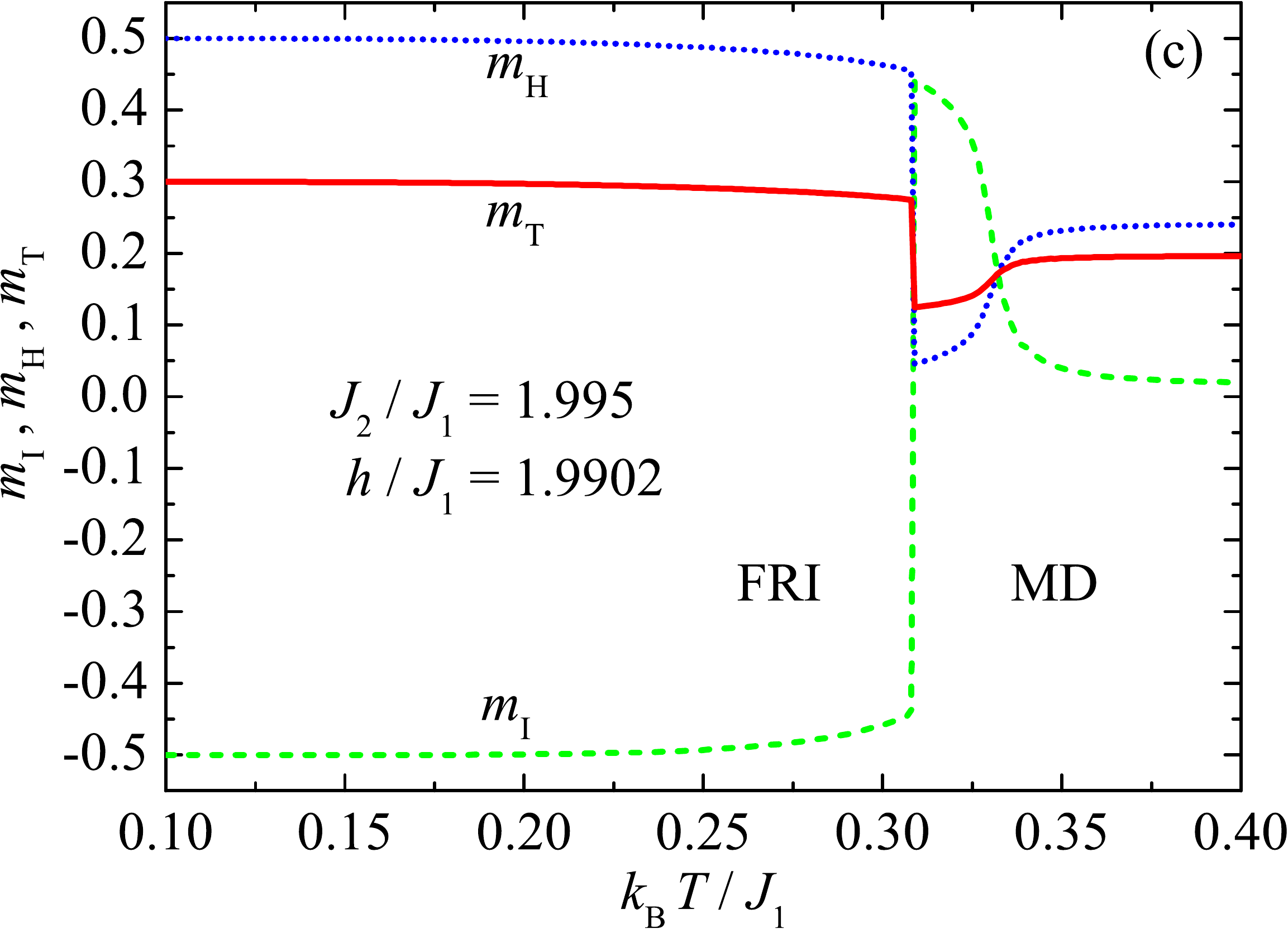}\hfill%
\includegraphics[width=\columnwidth]{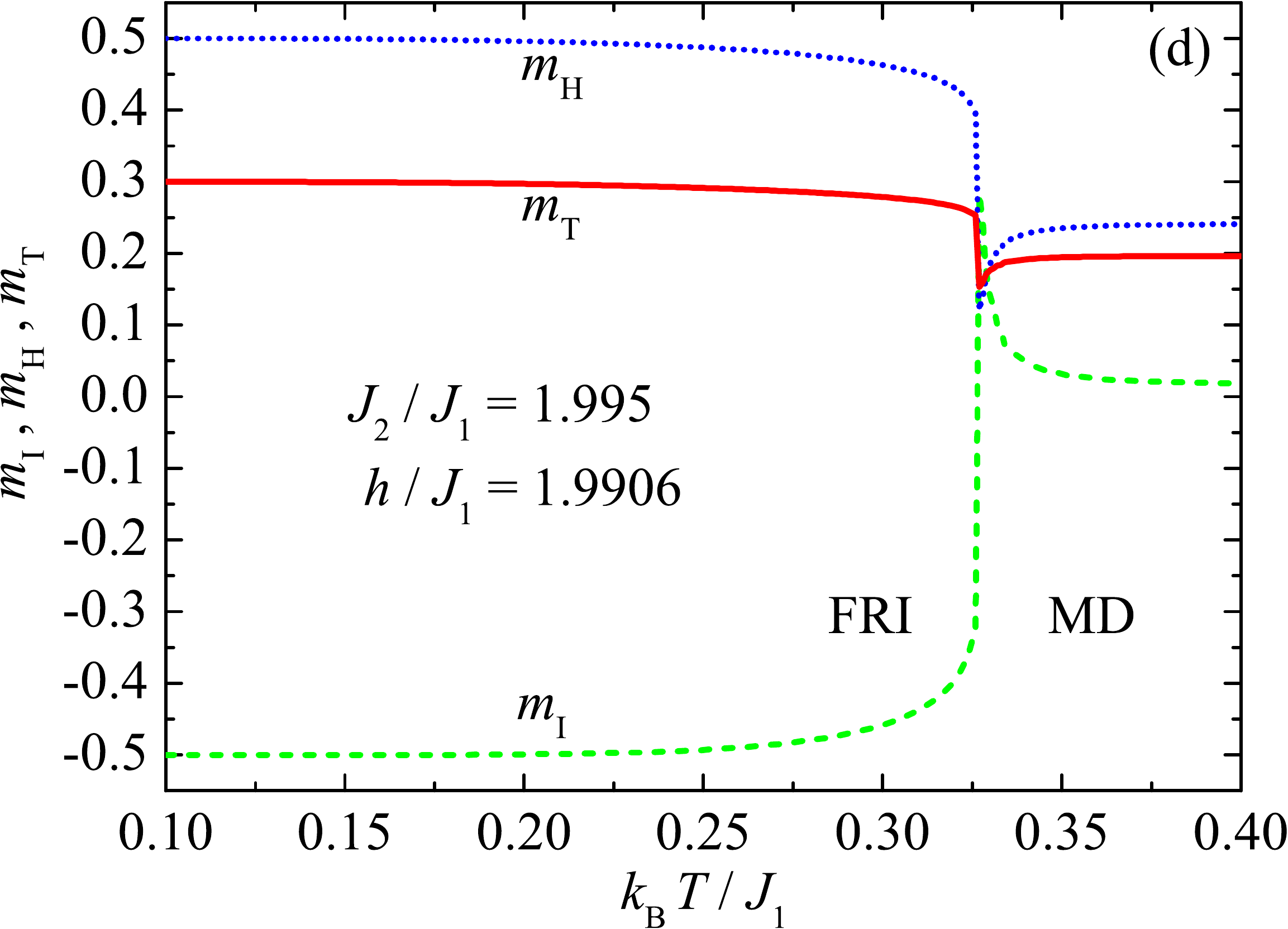}
\end{center}
\vspace{-5mm}
\caption{Results from Monte Carlo simulations for the temperature dependencies of the magnetization of the spin-1/2 Ising-Heisenberg model on the diamond-decorated square lattice for a fixed value of the interaction ratio $J_2/J_1 = 1.995$ and four different values of the magnetic field: (a) $h/J_1 = 1.9897$, (b) $h/J_1 = 1.9899$, (c) $h/J_1 = 1.9902$, (d) $h/J_1 = 1.9906$.}
\label{mcdtd}       
\end{figure*}

A few typical plots of the singlet density are shown in Fig.\ \ref{sdam2}(a) as a function of the magnetic field for a fixed value of the interaction ratio $J_2/J_1=2$ and four selected temperatures. At sufficiently low temperatures one indeed observes in Fig.\ \ref{sdam2}(a) an abrupt jump in the singlet density from a relatively high value $n \approx 1$ ascribed to the MD phase 
to a rather low value $n \approx 0$ corresponding to the FRI phase. The discontinuous jump in the singlet density coincides with the discontinuous field-induced phase transition, and gradually shrinks upon increasing of temperature until it completely vanishes at the critical temperature $k_{\rm B}T/J_1 \approx 0.33$. The continuous field-driven phase transition at the respective critical temperature is reflected in the singlet density through the Ising critical point denoting an inflection point with infinite tangent. Figure \ref{sdam2}(a) shows that the singlet density monotonically decreases with  increasing  magnetic field at higher temperatures, e.g., at  $k_{\rm B}T/J_1=0.4$. Finally, we have compiled in Fig.\ \ref{sdam2}(b) the total magnetization as a function of the magnetic field for a fixed value of the interaction ratio $J_2/J_1=2$ and four different values of temperature. This comparison bears evidence of a fundamental difference between finite-temperature manifestations of two discontinuous zero-temperature phase transitions, MD-FRI and FRI-PM, respectively. While the former discontinuous phase transition MD-FRI is still preserved at sufficiently low temperatures bounded from above by the Ising critical point, the latter phase transition, FRI-PM, merely exists at zero temperature and is replaced by simpler crossover  without any singularity for arbitrary nonzero temperatures.

\begin{figure*}
\begin{center}
\includegraphics[width=\columnwidth]{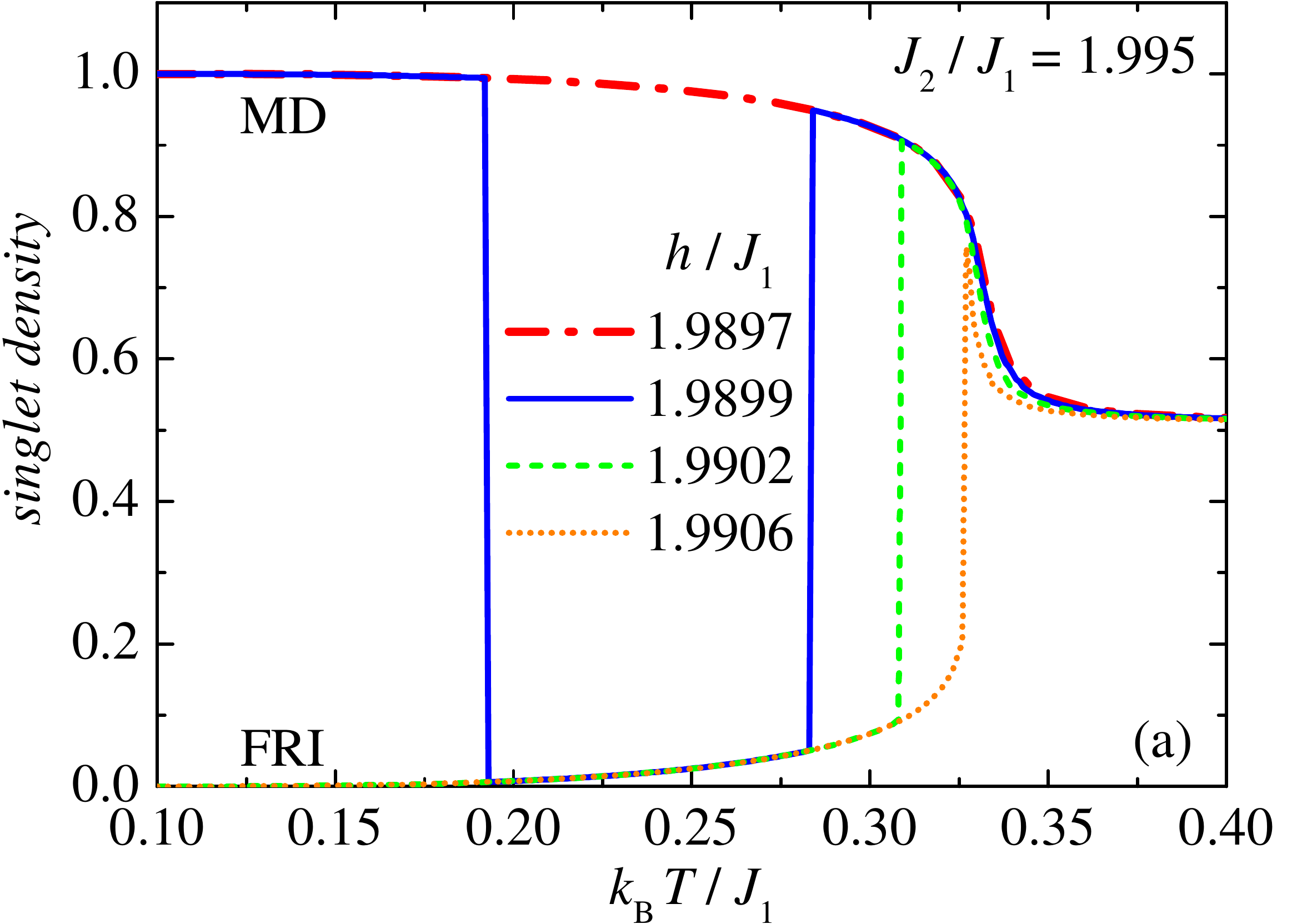}\hfill%
\includegraphics[width=\columnwidth]{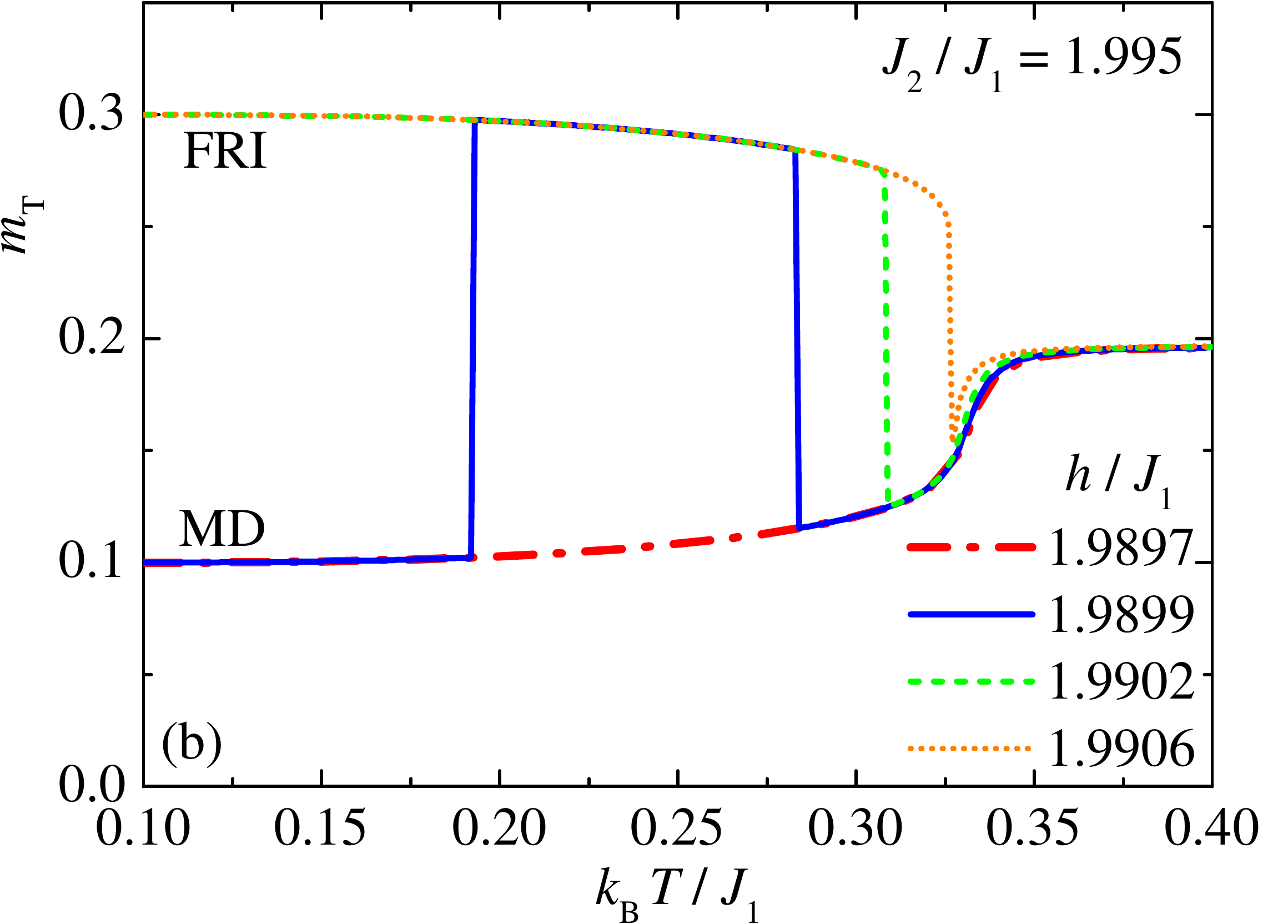}
\end{center}
\caption{Temperature dependence of the singlet density (a) and the total magnetization (b) of the spin-1/2 Ising-Heisenberg model on the diamond-decorated square lattice for a fixed value of the interaction ratio $J_2/J_1=1.995$ and four different values of the magnetic field.}
\label{sdr}       
\end{figure*}

Another 
question 
is whether Monte Carlo simulations can corroborate the reentrant phase transitions, which emerge only within a very narrow range of the magnetic fields, as exemplified 
in Fig.\ \ref{ftpd}(c) and (d). As a showcase, we depict in Fig.\ \ref{mcdtd} Monte Carlo data for the temperature dependencies of the magnetization of the spin-1/2 Ising-Heisenberg model on the diamond-decorated square lattice for a fixed value of the interaction ratio $J_2/J_1=1.995$ and four different values of the magnetic field. 
The Monte Carlo data for the magnetization turn out to be in excellent agreement with the exact results presented in Fig.\ \ref{ftpd}(d). If the magnetic field is fixed to the value $h/J_1=1.9897$, the local and total magnetization exhibit the MD ground state, whose spin arrangement gradually melts upon increasing of temperature, without passing through any finite-temperature phase transition [see Fig.\ \ref{mcdtd}(a)]. The most peculiar thermal variations in the magnetization, which document the double reentrant phase transitions, are displayed in Fig.\ \ref{mcdtd}(b) for a magnetic-field strength of $h/J_1=1.9899$. Although this value of the magnetic field differs from the previous one only 
in the fourth decimal place, the Monte Carlo data for the discontinuous magnetization jumps are in excellent accordance with the exact values of the transition temperatures. The MD ground state actually persists up to the first transition temperature $k_{\rm B}T/J_1\approx 0.1875$, at which the
system undergoes a phase transition to the FRI phase, which is then  stable until a reentrant phase transition at the second transition temperature $k_{\rm B}T/J_1\approx 0.285$, leading back into the MD phase. Note that the phenomenon of the double reentrance suddenly disappears by an additional small rise of the magnetic field to $h/J_1=1.9902$, for which one merely detects a single discontinuous phase transition from the FRI phase to the MD phase [see Fig.\ \ref{mcdtd}(c)]. Last but not least, temperature variations of the local and total magnetization at even higher magnetic field $h/J = 1.9906$ vary continuously as one reaches the Ising critical point, which marks a continuous phase transition between the FRI and MD phases [see Fig.\ \ref{mcdtd}(d)]. 

\begin{figure*}[t!]
\begin{center}
\includegraphics[width=\columnwidth]{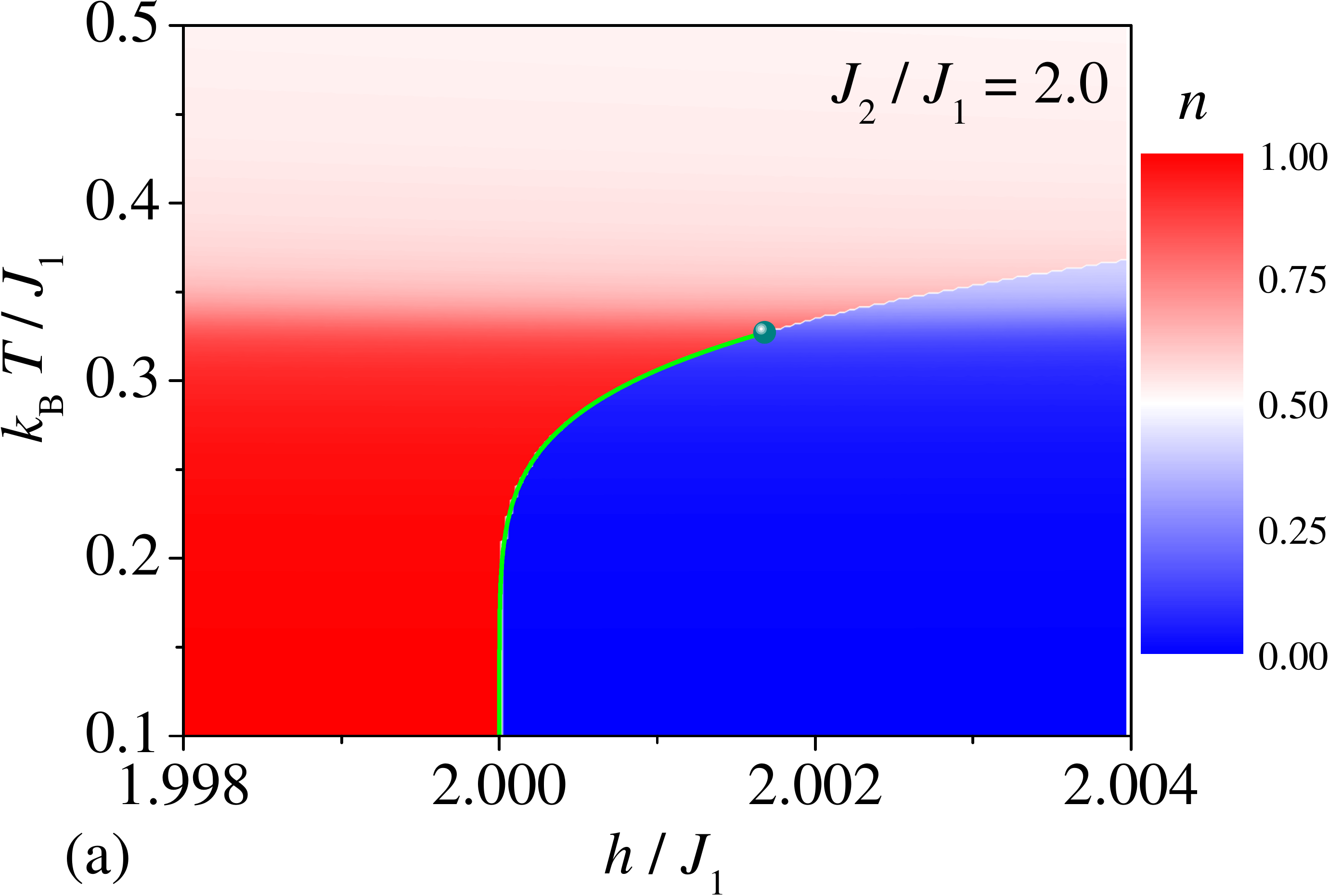}\hfill%
\includegraphics[width=\columnwidth]{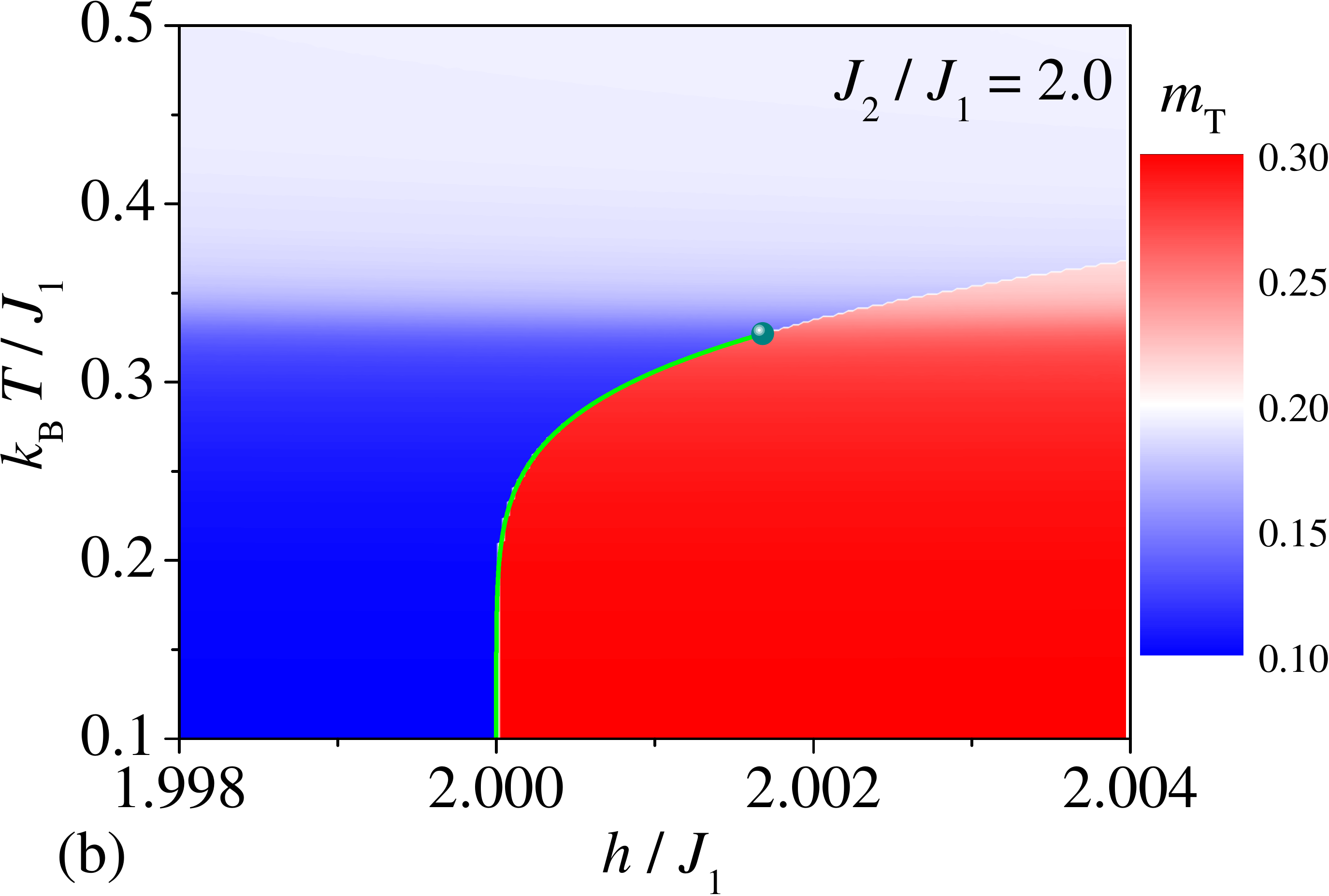} \\[2mm]
\includegraphics[width=\columnwidth]{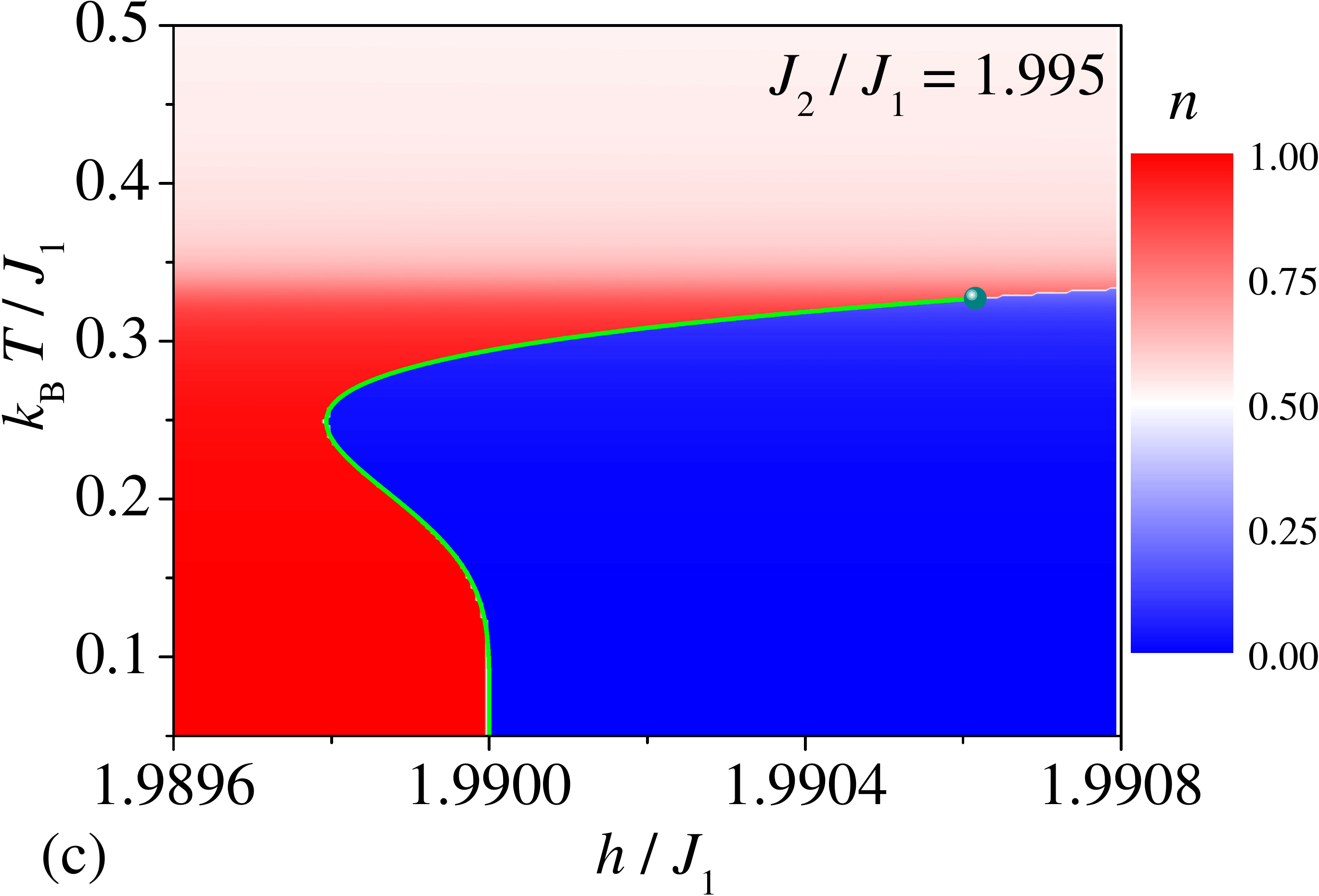}\hfill%
\includegraphics[width=\columnwidth]{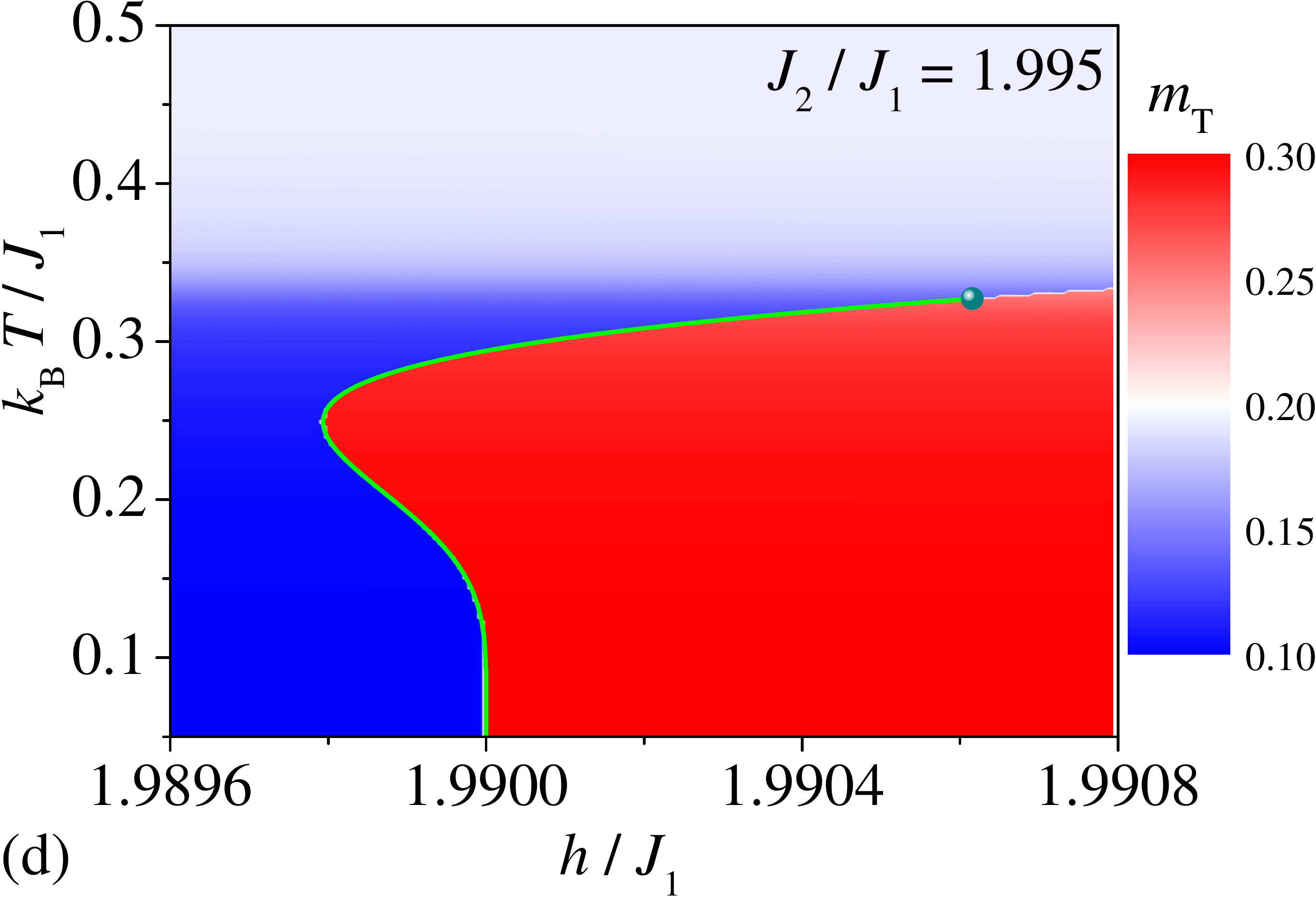}
\end{center}
\vspace{-5mm}
\caption{Monte Carlo results for  the singlet density $n$ (left panel) and the total magnetization $m_T$ (right panel) of the spin-1/2 Ising-Heisenberg model on the diamond-like decorated square lattice for two selected values of the interaction ratio: (a)--(b) $J_2/J_1=2.0$, (c)--(d) $J_2/J_1=1.995$. Thick green solid lines display exact results for the line of first-order phase transitions terminated by the Ising critical point (green ball).}
\label{densplots}      
\end{figure*}

Monte Carlo data for the temperature dependence of the singlet density are 
plotted in Fig.\ \ref{sdr}(a) for a fixed value of the interaction ratio $J_2/J_1=1.995$ and four different values of the magnetic field. At low  temperatures the singlet density reaches its maximal value for a magnetic field of $h/J=1.9897$, due to the presence of the MD phase, whose spin arrangement is subject to a gradual temperature-driven melting, mainly on account of a thermally-assisted activation of the FRI state. On the other hand, two discontinuous jumps in the singlet density are detected when the magnetic field is set to $h/J=1.9899$ as to achieve a sequence of double reentrant phase transitions MD-FRI and FRI-MD [compare Fig.\ \ref{sdr}(a) with the phase diagram shown in Fig.\ \ref{ftpd}(d)]. The two discontinuities in the singlet density can be alternatively regarded as a smoking gun evidence of the discontinuous reentrant phase transitions. For a slightly larger magnetic field, $h/J=1.9902$, the singlet density is zero at low temperatures, in agreement with the FRI ground state, and it gradually increases upon rising temperature until it shows a single discontinuous jump, indicating the first-order phase transition from the FRI phase to the MD phase. Finally, a continuous change in the singlet density is expected at the Ising critical point related to the second-order phase transition between the FRI  and  MD phase for the particular case with $h/J=1.9906$, shown in Fig.\ \ref{sdr}(a). 

To complete the overall picture we  plotted in Fig.\ \ref{sdr}(b) the total magnetization of the spin-1/2 Ising-Heisenberg model on the diamond-decorated square lattice against temperature for the same set of parameters as  used in Fig.\ \ref{sdr}(a).
The zero-temperature asymptotic value of the total magnetization $m_T=0.1$ (i.e., the 1/5-plateau) corresponds to the MD phase, while $m_T=0.3$ (i.e., the 3/5-plateau) is indicative of the FRI phase. Although the total magnetization never achieves the exact zero-temperature values of $m_T=0.1$ and $0.3$ at nonzero temperatures, the transition values of the total magnetization always fall into the relevant branches that are predicted for the pure MD phase (as obtained for instance for $h/J=1.9897$) or the pure FRI phase (as obtained for $h/J=1.9906$).
Furthermore, according to Fig.\ \ref{sdr}(b) the temperature dependencies of the total magnetization exhibit the following variants of the magnetic-field driven phase transitions, depending on the strength of the magnetic field: the absence of any phase transition ($h/J=1.9897$), the double reentrant discontinuous phase transitions manifested by two discontinuous magnetization jumps ($h/J=1.9899$), a single discontinuous phase transition evidenced by a unique magnetization jump ($h/J=1.9902$), or a single continuous phase transition through an inflection point with vertical tangent ($h/J=1.9906$). 

The preceding analysis shows that the singlet density and the total magnetization of the spin-1/2 Ising-Heisenberg model on the diamond-decorated square lattice exhibit  pronounced changes across the line of discontinuous phase transitions, because both  quantities take on rather distinct values within the MD and FRI phases. Based on this fact, the density plots of the singlet density $n$ and the total magnetization $m_T$, displayed in Fig.\ \ref{densplots} in the magnetic field versus temperature plane, could be alternatively regarded as 
finite-temperature phase diagrams.
Figure \ref{densplots} demonstrate that the singlet density and the total magnetization indeed undergo abrupt discontinuous changes along the whole first-order phase-transition lines displayed in Fig.\ \ref{densplots} by  thick green lines, whereas a smooth continuous change of these quantities can be observed at and above the Ising critical points, depicted by green spheres. The singlet density and the total magnetization are considerably smeared out above the Ising critical temperature, where they exhibit merely  subtle deviations from the mean values $n = 0.5$ and $m_T = 0.2$, respectively. Last but not least, 
the results of Monte Carlo simulations shown in Figs.\ \ref{densplots}(c)--(d) for the particular value of the interaction ratio $J_2/J_1=1.995$ confirm the previously predicted reentrant discontinuous phase transitions from the MD phase to the FRI phase and vice versa.

\section{Concluding remarks}
\label{conclusion}

We  investigated 
the magnetic properties of the spin-1/2 Ising-Heisenberg model on the  diamond-decorated square lattice in a magnetic field by  means of a generalized decoration-iteration transformation and classical Monte Carlo simulations. The generalized decoration-iteration transformation provided us with an exact mapping of the original model to an  effective spin-1/2 Ising model on the square lattice, with temperature-dependent effective nearest-neighbor interactions and magnetic field. 
Based on this mapping, the spin-1/2 Ising-Heisenberg model on the diamond-decorated square lattice in a magnetic field becomes exactly solvable within the  particular subspace of the parameter region where the effective field vanishes. Apart from a trivial case, one obtains a vanishing effective field along the ground-state boundary between the  FRI phase and the  MD phase, which gives rise to exactly determined lines of thermal first-order phase transitions, each terminating in a critical point.
The exact mapping shows that this critical point belongs to the Ising universality class.
We demonstrated that the first-order phase-transition lines bend  either towards lower or higher magnetic fields upon increasing  temperature, depending on the interaction ratio. This  bending of the phase-transition lines has been explained in terms of low-energy excitations atop the FRI and MD ground states. The remarkably good agreement that we demonstrate between the exact shapes of the first-order transition lines and their approximated form, obtained from the free-energy arguments put forward in Ref.~\onlinecite{sta18}, demonstrates explicitly the reliability of the latter approach, which has been used in various related studies~\cite{sta18,web22,web22b,cac22}. We note that Ref.~\onlinecite{web22b}  derived a mapping from the plaquettized fully frustrated bilayer spin-1/2 Heisenberg model onto an effective Ising model, which similarly describes a line of Ising critical points that terminates a wall of first-order transitions. However, those results were obtained only in perturbation theory, whereas here we obtained exact analytical results for the thermal transitions. 

Our exact results have furthermore corroborated the presence of reentrant discontinuous  phase transitions, which emerge inside a rather narrow but finite parameter regime. To further explore the parameter region in which the effective field is nonzero, we performed classical Monte Carlo simulations of the effective spin-1/2 Ising square lattice model with temperature-dependent interactions and magnetic field, from which we could extract accurate numerical results  for the spin-1/2 Ising-Heisenberg model on the diamond-decorated square lattice in a magnetic field.
These 
Monte Carlo results are in perfect agreement with the exact analytical results available for zero effective field and provided us with supplementary insights into the parameter regime with a nonzero effective field.  

We remark that similar lines of thermal first-order phase transitions, ending in Ising critical points, have been  reported recently also for the fully quantum spin-1/2 Heisenberg model on the diamond-decorated square lattice in a magnetic field, which has been treated by sign-problem-free quantum Monte Carlo simulations~\cite{cac22}. From this perspective, the exact analytical results for the simplified spin-1/2 Ising-Heisenberg model on the diamond-decorated square lattice in a magnetic field reported here are 
valuable as they bring about a more comprehensive understanding of several intriguing physical phenomena such as the bending of the first-order phase-transition lines, the presence of the Ising critical points as well as a reentrant discontinuous phase transition. Quantitatively accurate results for the full Heisenberg model require  time-consuming state-of-the-art numerical methods. Without  exact results, the reentrance phenomenon could for instance be easily overlooked, because this phenomenon emerges only within a rather narrow parameter region. It would certainly be interesting to investigate the possibility of similar reentrance phenomena also in the quantum spin-1/2 Heisenberg model on the diamond-decorated square lattice in a magnetic field.

The present article also paves the way towards finding exact first-order phase transitions in several other two-dimensional Ising-Heisenberg models in magnetic fields, which can be treated by the generalized algebraic mapping transformations \cite{fis59,roj09,str10}. The spin-1/2 Ising-Heisenberg model on the diamond-decorated Bethe lattice in a magnetic field represents one prototypical example of this type, where an abrupt jump in the low-temperature magnetization curve was previously observed, but  unfortunately remained unrelated to a thermal first-order phase transition (see Fig.\ 4(a)--(b) in Ref.~\onlinecite{eki12} and Fig.\ 6(c)--(d) in Ref.~\onlinecite{str13}). It will therefore be interesting to revisit the thermal physics of this model in future work, explicitly considering further thermodynamic quantities, such as the specific heat or the magnetic susceptibility, in the vicinity of the previously observed discontinuities.

\begin{acknowledgments}
This work was financially supported by the \v{S}tef\'anik programme for Slovak-France
bilateral projects under the contract No.\ SK-FR-19-0013 / 45125RC, by the grant of The
Ministry of Education, Science, Research and Sport of the Slovak Republic under the contract
No.\ VEGA 1/0105/20 and by the grant of the Slovak Research and Development Agency under the
contract No.\ APVV-20-0150. We 
acknowledge support by the Deutsche Forschungsgemein\-schaft
(DFG) through Grant No.\ WE/3649/4-2 of the FOR 1807 and RTG 1995. 
\end{acknowledgments}

\end{document}